\documentclass[twocolumn,final,aps,prl,showpacs,superscriptaddress,amsmath,amssymb,amsfonts,floatfix]{revtex4-1}

\usepackage{graphicx}
\usepackage{multirow}
\usepackage{epsfig}
\usepackage{txfonts}

\usepackage[normalem]{ulem}

\newcommand{\bra}[1]{\langle #1 |}
\newcommand{\ket}[1]{| #1 \rangle}

\newcommand{\RR}{\mathbf{R}}
\newcommand{\rr}{\mathbf{r}}
\newcommand{\kk}{\mathbf{k}}

\newcommand{\ii}{\mathrm{i}}
\newcommand{\ee}{e}
\newcommand{\GG}{\mathbf{G}}

\newcommand{\qq}{\mathbf{q}}

\newcommand{\pc}{c^{\vphantom{\dagger}}}

\newcommand{\pw}{w^{\vphantom{*}}}

\usepackage{url}
\usepackage{color}
\usepackage[colorlinks,breaklinks,bookmarks=true,citecolor=blue,linkcolor=red,urlcolor=blue]{hyperref}
\definecolor{darkred}{rgb}{0.7,0.0,0}

\begin{document}

\title{Evolution of the Coulomb interactions 
in correlated transition-metal perovskite oxides
from the constrained random phase approximation}

\author{Liang Si}
\email{siliang@nwu.edu.cn}
\affiliation{School of Physics, Northwest University, Xi'an 710127, China}
\affiliation{University of Vienna, Faculty of Physics and Center for Computational Materials Science, Kolingasse 14-16, A-1090, Vienna, Austria}

\author{Peitao Liu}
\email{ptliu@imr.ac.cn}
\affiliation{Shenyang National Laboratory for Materials Science, Institute of Metal Research, Chinese Academy of Sciences, 110016 Shenyang, Liaoning, China}

\author{Cesare Franchini}
\email{cesare.franchini@univie.ac.at}
\affiliation{University of Vienna, Faculty of Physics and Center for Computational Materials Science, Kolingasse 14-16, A-1090, Vienna, Austria}
\affiliation{Dipartimento di Fisica e Astronomia, Universit`a di Bologna, 40127 Bologna, Italy}

\date{\today}

\begin{abstract}

Determining the strength of electronic correlations of correlated electrons plays important roles in accurately describing the electronic structures and physical properties of transition-metal (TM) perovskite oxides . 
Here, we study the evolution of electronic interaction parameters as a function of $d$-electron occupancy in an extended class of TM perovskite oxides $AB$O$_3$ ($A$=Sr, Ca, and $B$=3$d$-5$d$ TM elements) using the constrained random-phase-approximation method adopting two distinct models: $t_{2g}$-$t_{2g}$ and $d$-$dp$.
For Sr$B$O$_3$ with $B$=Fe, Ru, and Ir, the $t_{2g}$-$t_{2g}$ model faces critical challenges, as the low-energy Hamiltonian spanning $t_{2g}$ manifolds is ill-defined. The $t_{2g}$-$t_{2g}$ model suggests that, for early $AB$O$_3$ series ($B$=$d^1$-$d^3$), the bare Coulomb interaction parameters $V$ remain nearly constant due to the competition between extended $t_{2g}$ Wannier orbitals and bandwidth reduction. As the $d$-electron filling increases, both partially screened Coulomb interaction parameters $U$ and fully screened Coulomb interaction parameters $W$ decrease, which are attributed to enhanced $e_g$-$t_{2g}$ and $e_g$-$p$ screenings. 
In contrast to the $t_{2g}$-$t_{2g}$ model, the $d$-$dp$ model effectively handles both early and late $AB$O$_3$ perovskites and reveals different trends. Specifically, $V$ varies inversely with the spreads of $d$-orbitals.
$W$ reaches its minimum at the $d^3$ occupancy due to an interplay between increasing $d$-orbital localization and increasing screening effects. An unusual trend is observed for $U$, with local maxima at both $d^1$ and $d^4$ occupations. This can be understood from two aspects: (1) the increasing full screening effects from $d^1$ to $d^3$ and (2) the strongest $d$-$d$ and the weakest $d$-$p$ screening effects near $d^4$ for Sr$B$O$_3$. Our results underscore that the Coulomb interaction parameters not only depend on the choice of model frameworks, but also the interplay between the localization of $d$-orbitals and screening strength, which are affected by $d$-orbital fillings and details of the band dispersion.

\end{abstract}

\maketitle
\section{I.\,Introduction}

The accurate simulations of material properties and phenomena through theoretical calculations rely heavily on understanding the intricate interactions between inhomogeneous distributed electrons. In many cases, these electronic interactions are governed by the crystal symmetry, lattices, and the constituent elements, making their precise determination challenging yet crucial for reliable theoretical predictions. For instance, after the experimental discovery and theoretical explanations for the Mott-insulator series \cite{fazekas1999lecture,mott1949basis}, the scientific community has been convinced by the importance of the physics behind electronic correlation effects. For example, the Hubbard $U$ explains the anti-ferromagnetic insulating state in transition-metal compounds \cite{PhysRevB.70.235121}, such as NiO \cite{PhysRevB.57.6884,PhysRevB.62.16392}, CeO$_2$ \cite{PhysRevB.75.035115}, Ce$_2$O$_3$ \cite{PhysRevB.75.035109}, CoO \cite{anisimov1991band} and cuprate superconductor CaCuO$_2$ \cite{anisimov1997first}. Additionally, the non-negligible $U$ is found to be the driving force behind the Mott-insulating transition in V$_2$O$_3$ \cite{PhysRevLett.86.5345}, and the orbital ordering state in cuprate KCuF$_3$ \cite{liechtenstein1995density}.  
Another noteworthy case is the exotic insulating state in Sr$_2$IrO$_4$ due to the interplay between  electronic correlation and spin-orbit interaction~\cite{PhysRevLett.101.076402,PhysRevLett.108.086403,PhysRevB.92.054428}. These observations demonstrate the significance of electronic correlation effects in oxides and their role in explaining emergent electronic and magnetic phases of matter.

Among the large family of oxide materials, transition-metal oxides (TMOs) perovskites are an important class due to the intricate interplay between various degrees of freedom, including correlations, charge, spin, orbital, and lattice~\cite{cox2010transition}. This complexity makes perovskite oxides a rich field of study for both experimental and theoretical research~\cite{tokura2000orbital}. Electronic correlation effects in TMOs perovskites are profound and impact virtually all their properties and potential functionalities.
For strongly electronic correlated systems, such as solids that contain partially filled transition-metal or rare-earth elements, standard density-functional theory (DFT) often fails to account for electronic and magnetic properties.
This necessitates the use of advanced beyond-DFT computational methods~\cite{PhysRev.136.B864,RevModPhys.61.689}, such as DFT+$U$~\cite{anisimov1991band,anisimov1997first}, dynamical mean-field theory (DMFT)~\cite{georges1996dynamical,kotliar2004strongly}, DFT+DMFT~\cite{PhysRevLett.62.324,held2008bandstructure,held2007electronic}, D$\Gamma$A~\cite{PhysRevB.75.045118,RevModPhys.90.025003} and GW+DMFT~\cite{PhysRevB.94.201106,PhysRevX.11.021006,held2011hedin,PhysRevB.95.155104,PhysRevB.66.085120,PhysRevLett.90.086402,Tomczak2017}. 
Theoretical research leveraging these advanced techniques has been pivotal in uncovering several exotic physical phenomena in perovskite oxides, including high-temperature superconductivity~\cite{PhysRevB.76.104509,PhysRevB.62.R9283}, colossal magnetoresistance~\cite{10.1063/1.110624,doi:10.1126/science.264.5157.413,Shimakawa1996GiantMI}, multiferroicity~\cite{Picozzi_2009,Zhao2014,VarignonBristoweBousquetGhosez+2020,bersuker2020perovskite,bersuker2022origin}, topological electronic states \cite{Chen2022,Irkhin2022} and exotic magnetic states~\cite{Kim2017_327,PhysRevB.94.241113,PhysRevLett.115.156401}. 
Among all these correlated methods, the inclusion of a suitable Hubbard $U$ is essential for obtaining a precise and materials specific description of the underlying physics.


However, accurately determining the electron interaction parameters is a nontrivial task. 
Physically, $U$ corresponds to the difference between the electron affinity and the ionization energy when respectively adding and removing an electron on the correlated shell (e.g., the $d$ shell) of a given atom, and in principle can be calculated from the experimental spectroscopy such as photoemission and inverse photoemission spectra~\cite{LDA+DMFT2011}. Nevertheless, this requires the presence of a natural localized basis in the system, which does not apply for a majority of correlated materials due to the orbitals hybridization or spin-orbit coupling effects.
The complexity of many-body interactions in solids often necessitates sophisticated theoretical frameworks and computational methodologies
such as many-body perturbation theory~\cite{bartlett1981many} or quantum Monte Carlo simulations~\cite{becca2017quantum}.
The fundamentally different formulation of the electronic structure problem in DFT and lattice Hubbard model introduces complications when attempting to precisely define the parameters $U$ and $J$ for the DFT++ methods. The situation is further exacerbated by the adoption of different basis functions in commonly employed DFT codes and advanced many-body methods.
In the past years, remarkable achievements have been made concerning the question of how to calculate $U$~\cite{PhysRevB.43.7570,PhysRevB.71.035105,TIMROV2022108455,PhysRevB.103.045141,PhysRevMaterials.8.014409}. Most often, the $U$ is taken as an adjustable parameter, which is obtained by fitting theoretical outputs with experimental spectra, e.g., X-ray photoemission and absorption spectra~\cite{PhysRevB.15.1669,Franchini2007}.
Within DFT, a common computational approach for obtaining $U$
is the constrained DFT [also referred to as constrained LDA (cLDA)] ~\cite{PhysRevB.43.7570,PhysRevB.71.035105}.
In this method, $U^{cLDA}$ for a given shell is defined as the derivative of the total energy with respect to the variation of the electron occupancy. 
The cLDA method has been implemented within different basis functions and  electronic-structure codes, e.g., in the (L)APW+lo framework~\cite{PhysRevB.64.195134} or in the basis of maximally localized Wannier functions~\cite{marzari2012maximally}, and applied to obtaining the $U$ value for various materials~\cite{PhysRevB.71.045103,PhysRevB.97.035117,PhysRevB.105.195153,PhysRevB.39.9028}.

An alternative widely adopted approach is the so-called constrained random phase approximation (cRPA), firstly proposed by Aryasetiawan \emph{et al.}~\cite{miyake2008screened,PhysRevB.70.195104}. This method is based on the
intuitive idea that the Hubbard $U$ can be viewed as a Coulomb interaction screened by the polarization of the whole
system excluding the polarization arising from a set of bands which are treated in the Hubbard model. 
For practical materials, the starting point is to downfold the full electronic structure of correlated systems to an effective low-energy Hamiltonian spanning by the correlated orbitals.
Then, the Hubbard $U$ and Hund's $J$ for the effective low-energy Hamiltonian are defined as the matrix elements of the effective partially screened interactions, excluding the screenings from the correlated subspace. The cRPA method has been implemented in various codes such as \textsc{Wien2k}~\cite{PhysRevB.86.165105},
\textsc{ABINIT}~\cite{PhysRevB.89.125110},
\textsc{FLEUR}~\cite{PhysRevB.83.121101},
\textsc{Tokyo Ab initio Program Package}~\cite{PhysRevB.86.085117},
\textsc{COULOMBU}~\cite{PhysRevB.85.045132},
SPEX~\cite{PhysRevLett.109.146401},
\textsc{VASP}~\cite{Merzuk2015},
and 
Quantum
Espresso ~\cite{doi:10.1021/acs.jctc.4c00085}.
The robust implementations therefore lead to wide applications of the cRPA method in computing the Hubbard $U$ interactions for a variety of materials families, e.g.,  
oxypnictides, parent compounds of Fe-based superconductors, and 3$d$-5$d$ transition metals, 4$f$ lanthanides, and their oxides~\cite{PhysRevB.83.121101,PhysRevB.74.125106,PhysRevB.80.155134,PhysRevMaterials.2.075003,PhysRevB.95.115111,PhysRevB.94.241113,Hampel2019,PhysRevResearch.5.L012008, PhysRevB.94.195145,PhysRevB.95.024406,PhysRevMaterials.5.035404,PhysRevB.104.035102,PhysRevB.103.195101,10.1063/5.0137264,PhysRevResearch.3.023027,PhysRevB.83.121101,PhysRevB.100.115113,PhysRevB.98.205101,Martins_2017,PhysRevB.86.165124,PhysRevB.71.045103,PhysRevB.97.035117,PhysRevB.96.045137,PhysRevB.89.125110,PhysRevB.104.045134,PhysRevB.98.235151,PhysRevB.71.035105}. 

Despite the numerous applications of the cRPA method, a comprehensive investigation on the evolution of electronic interaction parameters with the electron occupancy and bandwidth of the $d$-shell of TM perovskite oxides across the entire range of TM elements is currently still lacking. In particular, the trends delivered by the widely adopted two schemes (i,e., $t_{2g}$-$t_{2g}$ and $d$-$dp$) have not yet been carefully examined. The present work aims to fill in this gap by computing the bare Coulomb interaction parameters $V$,
partially screened Coulomb interaction parameters $U$ and fully screened Coulomb interaction parameters $W$ of a series of $AB$O$_3$ TM perovskites oxides ($A$=Sr and Ca, $B$=3$d$ TMs from V to Co, 4$d$ TMs from Nb to Rh, and 5$d$ TMs from Ta to Pt). The resulting trends over the electron occupancy and bandwidth of the $d$-shell within the $t_{2g}$-$t_{2g}$ and $d$-$dp$ approximations were revealed and discussed. This work underlines the significance of the choosing an appropriate model and considering the counterbalanced effects between the localization of $d$-orbitals and the strength of screenings in order to obtain a meaningful $U$ value, and thus serves as an important addition to the community of correlated electrons.


\section{II.\,Method and Computational Details}

\subsection{A. Brief overview of the cRPA method}\label{sec:overview}

Initially, the cRPA method~\cite{PhysRevB.70.195104, PhysRevB.74.125106}
was designed for calculating the Hubbard $U$ value to merge DFT with
dynamical mean field theory (DMFT)~\cite{PhysRevLett.90.086402,PhysRevLett.62.324,RevModPhys.68.13,RevModPhys.78.865,PSSB:PSSB200642053,Held_2008,LDA+DMFT}.
In this regard, we formulate the cRPA starting from the model Hamiltonian
\begin{equation}\label{eq:ham}
H=\sum\limits_{\alpha\beta\RR}t_{\alpha\beta}^{(\RR)}
c^\dagger_{\alpha\RR}\pc_{\beta\RR}+
\sum\limits_{\alpha\beta\gamma\delta\RR}U_{\alpha\beta \gamma\delta}
c^\dagger_{\alpha\RR} c^\dagger_{\beta\RR} \pc_{\delta\RR} \pc_{\gamma\RR},
\end{equation}
where $t_{\alpha\beta}^{(\RR)}$ describes the electron hopping
from the Wannier orbital $\ket{w_\alpha}$ to orbital $\ket{w_\beta}$ at the lattice vector $\RR$.
These hopping matrix elements can be obtained from
Wannier interpolation of the Kohn-Sham eigenvalues $\epsilon_{n\kk}$~\cite{LDA+DMFT}.
$U_{\alpha\beta\gamma\delta}$ describes the effective on-site interaction between two particles and
can be expressed as the expectation value of the partially screened Coulomb kernel $\mathcal{U}$~\cite{PhysRevB.86.165105,app11062527}
\begin{equation}\label{eq:U}
\begin{split}
U_{\alpha\beta\gamma\delta}
& =\lim\limits_{\omega\to0}\bra{\pw_\alpha,
\pw_\beta}\mathcal{U}(\rr, \rr', \omega)\ket{\pw_{\delta},\pw_{\gamma}} \\
& = \lim\limits_{\omega\to0}\iint d\rr d\rr' w^*_\alpha(\rr) w^*_\beta(\rr') \mathcal{U}(\rr,\rr',\omega) w_\delta(\rr) w_\gamma(\rr').
\end{split}
\end{equation}
Within the framework of the cRPA, $\mathcal{U}$
is calculated via the RPA but using a ``constrained" polarizability, i.e., the rest polarizability $\chi^r$~\cite{PhysRevB.70.195104,PhysRevB.74.125106}
\begin{eqnarray}\label{eq:Ukernel}
\mathcal{U} =\mathcal{V}+\mathcal{V}\chi^r \mathcal{U} \quad \Leftrightarrow \quad \mathcal{U}^{-1}=\mathcal{V}^{-1}-\chi^r.
\end{eqnarray}
Here, $\mathcal{V}$ is the bare Coulomb kernel.
The rest polarizability $\chi^r$ contains all the RPA polarization effects except those within the correlated space~\cite{PhysRevB.70.195104,PhysRevB.74.125106}
\begin{equation}\label{eq:splitting}
\chi^r = \chi - \chi^c,
\end{equation}
where $\chi$ is the total independent-particle polarizability
at the RPA level~\cite{PhysRev.82.625,PhysRev.85.338,PhysRev.92.609,PhysRev.106.364,PhysRev.111.442}
and $\chi^c$ is the correlated polarizability that contains the polarization effects within the correlated space only.
The removal of $\chi^c$ from $\chi$ is to avoid double counting,
since the polarization effects between the correlated electrons
have already been accounted for by the many-body methods such as DMFT~\cite{Tomczak2017},
thereby recovering the fully screened interaction $\mathcal{W}$ by
\begin{eqnarray}\label{eq:W}
\mathcal{W} =\mathcal{U}+\mathcal{U}\chi^c \mathcal{W} \quad \Leftrightarrow \quad \mathcal{W}^{-1}=\mathcal{U}^{-1}-\chi^c.
\end{eqnarray}

\subsection{B. Practical implementation of the cRPA method}\label{sec:overview}

As introduced above, the key part of the cRPA method is to
remove the contribution of the correlated states from the total polarizability
when calculating the partially screened Coulomb kernel.
Such removal can be straightforwardly conducted
either in the plane wave basis~\cite{PhysRevB.74.125106}
or in the Wannier function basis~\cite{PhysRevB.82.045105,PhysRevB.86.085117},
if  the correlated states form an isolated manifold around the Fermi level.
However, when the correlated states are entangled with those non-correlated (usually $s$ or $p$) states of the system,
the correlated space becomes not trivially defined and therefore the evaluation of $\chi^c$ has to be treated with care.

To address this problem, several methods have been proposed.
Miyake, Aryasetiawan, and Imada~\cite{PhysRevB.80.155134} proposed the disentanglement method.
In this method, the correlated space $\mathcal{C}$ is disentangled from the full Fock space
by diagonalizing the Hamiltonian in $\mathcal{C}$ and the remaining Fock space separately.
This yields a minimal basis set within a given energy window that spans only the correlated space.
The correlated polarizability is then obtained using the disentangled band structures based on the Alder and Wiser formula~\cite{PhysRev.126.413,PhysRev.129.62}.
It is worth noting that the disentanglement method suffers from the deficiencies that
it alters the band structure and the minimal basis set as
well as the resulting $U$ depend strongly on the chosen energy window of the Wannier functions~\cite{PhysRevB.80.155134}.

To weaken the drawback of the disentanglement method,
Sasioglu, Friedrich and Bl\"{u}gel~\cite{PhysRevB.83.121101} proposed the weighted method.
In this method, non-correlated delocalized $s$- and/or $p$-states are included in the Wannier projection
and the effective interaction is calculated using a weighted polarizability,
where the weights are defined as probabilities for the Bloch states being correlated~\cite{PhysRevB.89.125110, PhysRevB.83.121101,PhysRevB.85.045132}.
Although the weighted method does not change the band structure,
it neglects the contributions to the polarizability from the non-diagonal terms of the correlated projectors~\cite{Merzuk2015}.

A more consistent and elegant method for computing the correlated polarizability is the projector method,
which was proposed by Kaltak~\cite{Merzuk2015} based on the Kubo--Nakano formula.
This is also the method employed in the present work. In the following, we briefly summarize the key parts of this method.
For more detailed deviations, we refer to Refs.~\cite{Merzuk2015,10.25365/thesis.48831}.
Within the projector method, the correlated polarizability in reciprocal space is calculated as~\cite{PhysRev.126.413,PhysRev.129.62,Merzuk2015}
\begin{equation}\label{eq:chic}
\begin{split}
\chi_{\GG\GG'}^c(\qq,\omega)
= & \frac{1}{N_{\kk}}\sum\limits_{\kk,n,n'}
\frac{f_{n\kk}-f_{n'\kk-\qq}}{\omega+\epsilon_{n\kk}-\epsilon_{n'\kk-\qq}-\ii\,\eta\,{\rm sgn}(\epsilon_{n\kk}-\epsilon_{n'\kk-\qq})} \\
\times\,
& \bra{  \bar\psi_{n\kk} }\ee^{\ii(\qq+\GG)\rr}\ket{  \bar\psi_{n'\kk-\qq} }
\bra{ \bar\psi_{n'\kk-\qq} }\ee^{-\ii(\qq+\GG')\rr'}\ket{ \bar\psi_{n\kk} }.
\end{split}
\end{equation}
Here, $\epsilon_{n\kk}$ are the Kohn-Sham eigenvalues, $f_{n\kk}$ are the occupancies and $\eta$ is a positive infinitesimal.
$\ket{\bar\psi_{n\kk}}$ is the projected Bloch function, which is defined as~\cite{Merzuk2015}
\begin{equation}\label{eq:function_correlated}
\ket{\bar\psi_{n\kk}}=\sum\limits_{m} P_{mn}^{(\kk)}\ket{\psi_{m\kk}},
\end{equation}
where $\ket{\psi_{n\kk}}$ are the Kohn-Sham eigenstates
and $P_{mn}^{(\kk)}$ is the projection matrix with the correlated projector defined as~\cite{Merzuk2015}
\begin{equation}\label{eq:projector1}
\hat{P}^{(\kk)}=\sum\limits_{\alpha\in\mathcal{C}}\ket{\Psi_{\alpha \kk}}\bra{\Psi_{\alpha \kk}}.
\end{equation}
Here, the summation is restricted to the correlated space $\mathcal{C}$
and $\ket{\Psi_{\alpha \kk}}$ is the mixed sate defined as~\cite{Merzuk2015}
\begin{equation}\label{eq:projector2}
\ket{\Psi_{\alpha \kk}}=\sum\limits_n T_{n \alpha}^{(\kk)} \ket{\psi_{n\kk}},
\end{equation}
where $T_{n \alpha}^{(\kk)}$ is the unitary matrix used to project the Bloch functions to the Wannier functions
and can be obtained from the wannier90 code~\cite{PhysRevB.56.12847,PhysRevB.81.054434,Mostofi2008685,CF1}.
Using Eq.~\eqref{eq:projector1} and Eq.~\eqref{eq:projector2} one can obtain the compact form of the projection matrix as
\begin{equation}\label{eq:projector_matrix}
P_{mn}^{(\kk)}=\bra{\psi_{m\kk}}\hat{P}^{(\kk)}\ket{\psi_{n\kk}}
=\sum\limits_{\alpha\in\mathcal{C}}T_{m \alpha }^{(\kk)}T_{n \alpha }^{*(\kk)}.
\end{equation}

Having determined the correlated polarizability [Eq.~\eqref{eq:chic}],
the rest polarizability $\chi^r$ in reciprocal space is then obtained using Eq.~\eqref{eq:splitting}
and the partially screened Coulomb kernel $\mathcal{U}_{\GG\GG'}(\qq,\omega)$
for every $k$-point $\qq$ in the irreducible wedge of the Brillouin zone is obtained using Eq.~\eqref{eq:Ukernel}.
Finally, the effective interaction matrix is evaluated via Eq.~\eqref{eq:U}.
After some mathematical derivations~\cite{Merzuk2015,10.25365/thesis.48831},
the explicit effective interaction matrix is given by
\begin{equation}\label{eq:matrix}
\begin{split}
U_{\alpha\beta\gamma\delta}(\omega)
=&
\frac{1}{N_{\qq}N^2_{\kk}}\sum\limits_{\qq,\kk,\kk'}\sum\limits_{\GG,\GG'}
\mathcal{U}_{\GG\GG'}(\qq,\omega) \\
\times\,&
\bra{\Psi_{\alpha \kk}}\ee^{\ii(\qq+\GG)\rr}\ket{\Psi_{\delta\kk-\qq}}
\bra{\Psi_{\beta\kk'-\qq}}\ee^{-\ii(\qq+\GG')\rr'}\ket{\Psi_{\gamma\kk'}}.
\end{split}
\end{equation}
In deriving above formula, we have used the translation-invariant symmetry and Fourier transformation
\begin{equation}\label{eq:matrix1}
\begin{split}
\mathcal{U}(\rr, \rr', \omega) =
\frac{1}{N_{\qq}}\sum\limits_{\qq}\sum\limits_{\GG,\GG'} \ee^{\ii(\qq+\GG)\rr}\mathcal{U}_{\GG\GG'}(\qq,\omega) \ee^{-\ii(\qq+\GG')\rr'},
\end{split}
\end{equation}
as well as the definition of the Wannier functions
\begin{equation}\label{eq:transformation22}
\ket{w_{\alpha}}\equiv \ket{w_{\alpha\RR}}=\frac{1}{N_{\kk}}\sum\limits_{\kk}\ee^{-\ii\kk\RR} \ket{\Psi_{\alpha \kk}}.
\end{equation}

From the full effective interaction matrix [Eq.~\eqref{eq:matrix}]
one can obtain the so-called Hubbard-Kanamori
parameters (intra-orbital interaction $U$, inter-orbital interaction $U'$ and Hund's coupling $J$)~\cite{PhysRevB.86.165105,Kanamori1963} by
\begin{eqnarray}
\label{eq:Udef}
U&=&\frac{1}{N}\sum\limits_{\alpha}^{N}U_{\alpha\alpha\alpha\alpha}(\omega=0),\\
\label{eq:Updef}
U'&=&\frac{1}{N(N-1)}\sum\limits_{\alpha\neq\beta}^{N}U_{\alpha\beta\beta\alpha}(\omega=0),\\
\label{eq:Jdef}
J&=&\frac{1}{N(N-1)}\sum\limits_{\alpha\neq\beta}^{N}U_{\alpha\beta\alpha\beta}(\omega=0).
\end{eqnarray}
Here, $N$ is the number of correlated states that span the correlated space.
Similarly, the bare and fully screened Coulomb interactions can be calculated as
\begin{eqnarray}
\label{eq:V}
V&=&\frac{1}{N}\sum\limits_{\alpha}^NV_{\alpha\alpha\alpha\alpha}(\omega=0), \\
\label{eq:W}
W&=&\frac{1}{N}\sum\limits_{\alpha}^NW_{\alpha\alpha\alpha\alpha}(\omega=0),
\end{eqnarray}
where the matrix elements $V_{\alpha\beta\gamma\delta}$ and
$W_{\alpha\beta\gamma\delta}$ are obtained using Eq.~\eqref{eq:matrix}
but with $\mathcal{U}$ replaced with $\mathcal{V}$ and $\mathcal{W}$, respectively.

\subsection{C. Computational details}

First-principles DFT and cRPA calculations were conducted using the \textsc{VASP} code~\cite{PhysRevB.47.558, PhysRevB.54.11169}. 
The exchange-correlation functional parameterized by Perdew-Burke-Ernzerhof (PBE) was employed~\cite{perdew1996generalized}. 
The energy cutoffs for the plane-wave basis set  and the response function were set to 500\,eV and 333\,eV, respectively.
Structural relaxation using the GGA-PBE correlation-exchange functional \cite{perdew1996generalized} often results in overestimated lattice constants and volumes. To address this, a modified version of PBE, known as PBE for solids (PBEsol), was proposed \cite{PhysRevLett.100.136406}, which provides better agreement between theoretical relaxed lattices and experimental values. However, the slightly modified lattices result in negligible changes in the $V$, $U$ and $J$ parameters. Therefore, we continue to use the standard PBE for relaxations and calculations.
A $\Gamma$-centered $k$-point grid 13$\times$13$\times$13 was employed for sampling the Brillouin zone.
The Gaussian smearing method with a smearing width of 0.05\,eV was used. 
The lattice constants for all the $AB$O$_3$ compounds considered were optimized before performing the electronic structure, Wannier projections, and cRPA calculations.
The convergence criteria for the electronic optimization and structural relaxation were set to 10$^{-8}$\,eV and  1\,meV/$\AA$, respectively.
To construct the low-energy Hamiltonian and obtain the projection matrix, 
the maximally localized Wannier functions within the $B$-$t_{2g}$ or $B$-$d$+O-$2p$ subspace were conducted 
using the \textsc{Wannier90} code~\cite{PhysRevB.56.12847,PhysRevB.81.054434,Mostofi2008685}.
The effective interaction matrix were calculated within the cRPA method using the projector method~\cite{Merzuk2015}.

\section{III. Results and Discussion}

\subsection{A. DFT band structure}

We start our discussion by presenting the band structures of $AB$O$_3$ perovskites. 
Since the trends over the $d$-electron filling of TM atoms are the main focus of our study, we limit ourselves to the (undistorted) cubic phase of $AB$O$_3$ perovskite TMOs. The cubic phase with $Pm3m$ space group was considered as the ground-state structure for some $AB$O$_3$ perovskites, such as 3$d$ perovskite SrVO$_3$ \cite{PhysRevB.43.181} and SrCrO$_3$ \cite{PhysRevB.84.125114}, but not for some 4$d$ and 5$d$ $AB$O$_3$ compounds, such as SrNbO$_3$ \cite{hannerz1999transmission}, SrRuO$_3$ \cite{fukunaga1994magnetism} and SrIrO$_3$ \cite{fruchter2019growth}. Here, we employ the cubic $Pm3m$ structure for all $AB$O$_3$ compounds to calculate the band structures and interaction parameters, as this simplification significantly reduces the computational efforts. With this setup, the TM cations in $AB$O$_3$ are octahedrally coordinated with O ligands, leading to band splittings between $d$-orbitals into threefold degenerate $t_{2g}$ and double degenerate $e_g$ states. Compared with $t_{2g}$ orbitals, the $e_g$ orbitals host stronger $d$-$p$ hybridization because their orbitals' lobes directly point to the O-$p$ orbitals along $x$-, $y$- and $z$-directions. The atomic number of the $B$-site changes the interaction parameters by variation of the lattice constants and the $d$-band filling. Based on the relaxed lattice constants collected in Table~\ref{table1}, the relaxed lattice in 3$d$-5$d$ $AB$O$_3$ decreases as the atomic number of $B$ increases. Hence, the $d$-band filling is expected to play a more important role. Additionally, as the $d$-band filling increases, the shift of Fermi energy leads to smaller energetic separation between $d$-band and O-$2p$ band, thereby further inducing strong $d$-$p$ hybridization in addition to the lattice shrinking.

\begin{figure*}
\centering
\includegraphics[width=18.0cm]{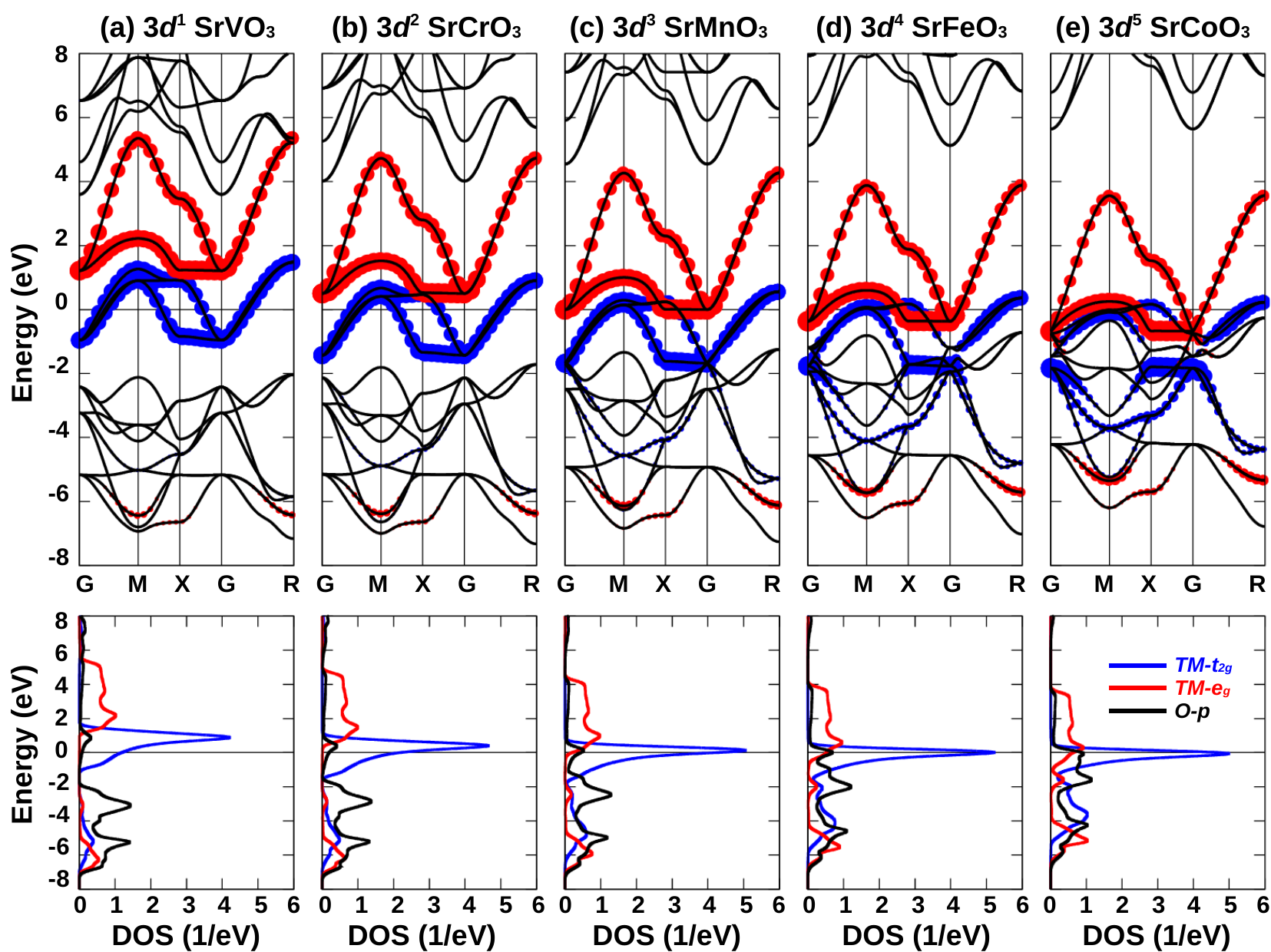}
\caption{(Color online) 
DFT bands (top panels) and density of states (bottom panels)
of Sr$B$O$_3$ ($B$=V, Cr, Mn, Fe, and Co).
The size of the blue and red points indicates contributions from $t_{2g}$ and $e_g$ orbitals.}
\label{Fig-Srband1}
\end{figure*}

\begin{figure*}
\centering
\includegraphics[width=18.0cm]{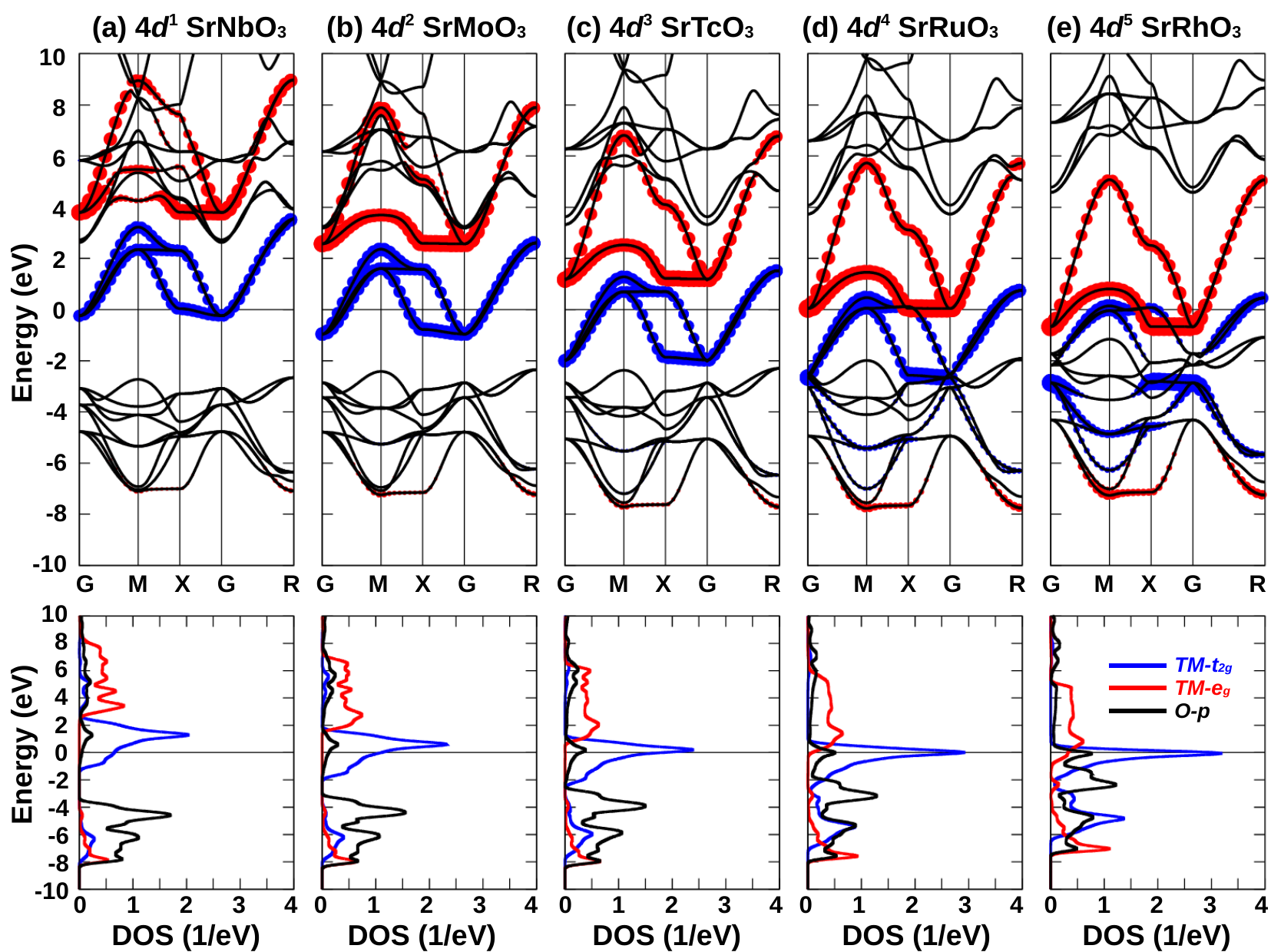}
\caption{(Color online) 
DFT bands (top panels) and density of states (bottom panels)
of Sr$B$O$_3$ ($B$=Nb, Mo, Tc, Ru, and Rh).
The size of the blue and red points indicates contributions from $t_{2g}$ and $e_g$ orbitals.
}
\label{Fig-Srband2}
\end{figure*}

\begin{figure*}
\centering
\includegraphics[width=18.0cm]{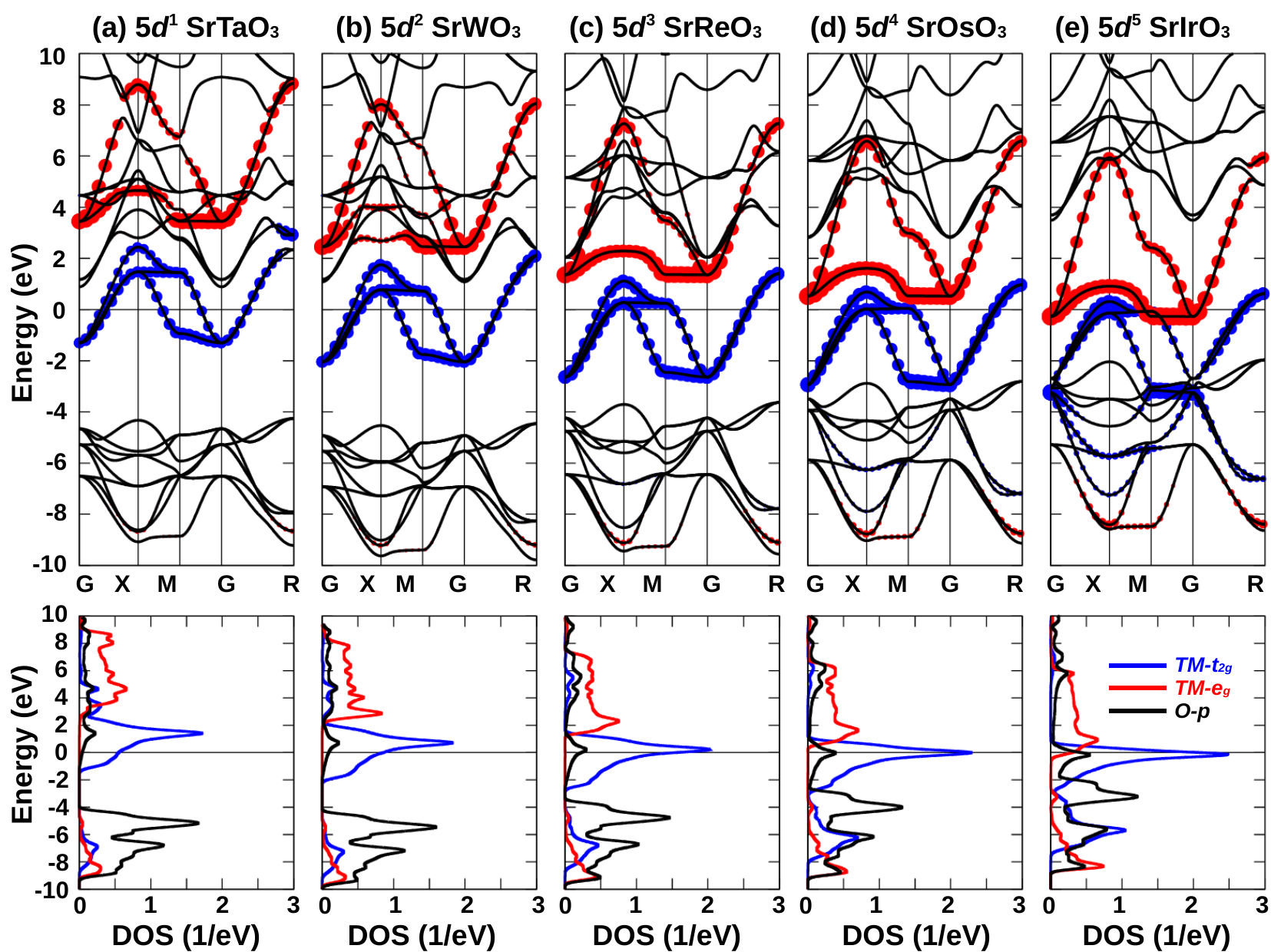}
\caption{(Color online) 
DFT bands (top panels) and density of states (bottom panels)
of Sr$B$O$_3$ ($B$=Ta, W, Re, Os, and Ir).
The size of the blue and red points indicates contributions from $t_{2g}$ and $e_g$ orbitals.}
\label{Fig-Srband3}
\end{figure*}

\begin{figure*}
\centering
\includegraphics[width=18.0cm]{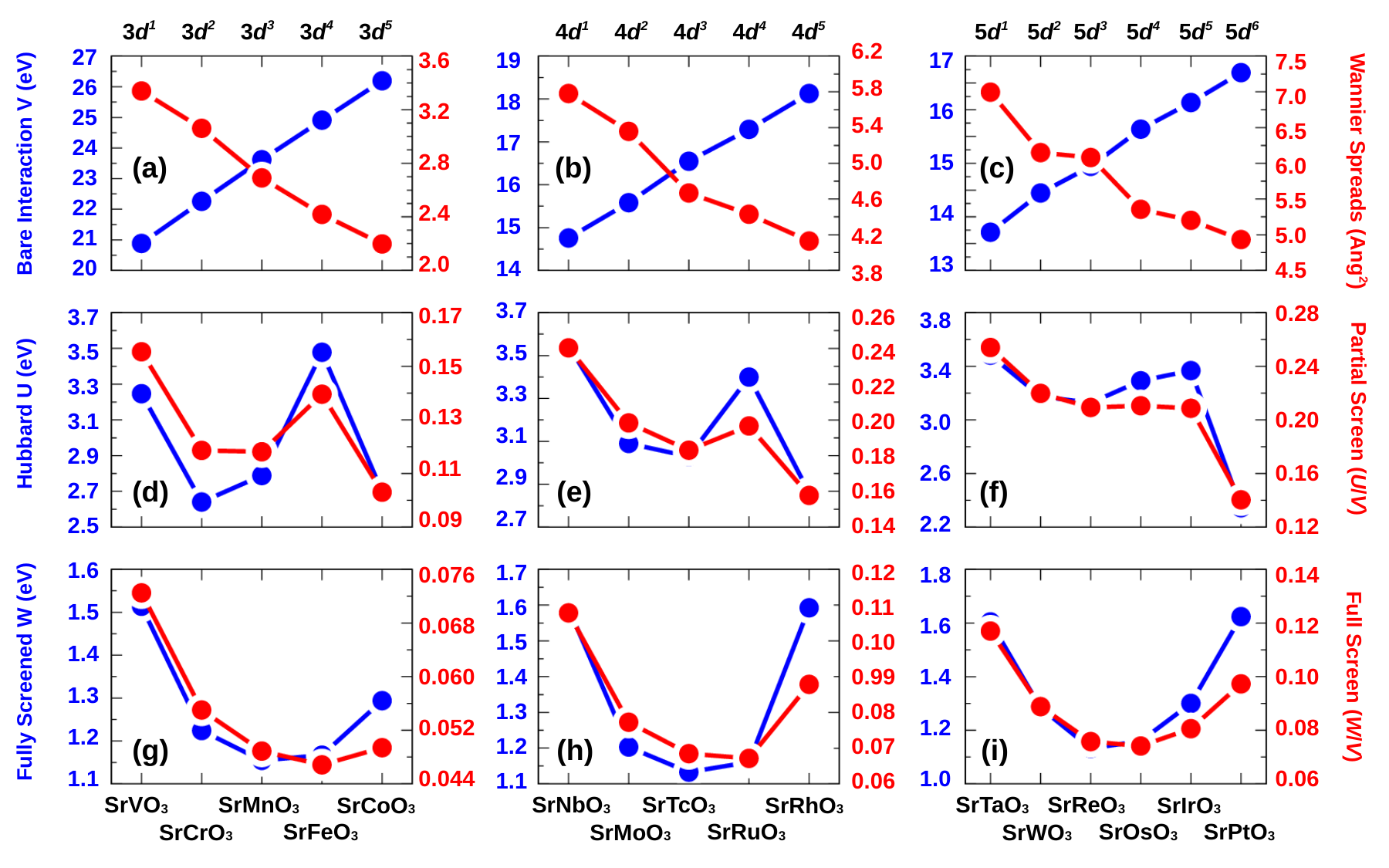}
\caption{The cRPA obtained interaction parameters for Sr$B$O$_3$ ($B$=V to Fe, Nb to Ru, and Ta to Os) within $d$-$dp$ approximation. Top panels: bare interaction $V$ (blue) vs. Spreads (red) of $d$-orbitals. Middle panels: Hubbard interaction $U$ (blue) (partially screened interaction) vs. partial screening (red) ($U$/$V$). Bottom panels: fully screened interaction $W$ (blue) vs. full screening (red) ($W$/$V$).}
\label{Fig-SrUJV-dp}
\end{figure*}

\begin{figure*}
\centering
\includegraphics[width=18.0cm]{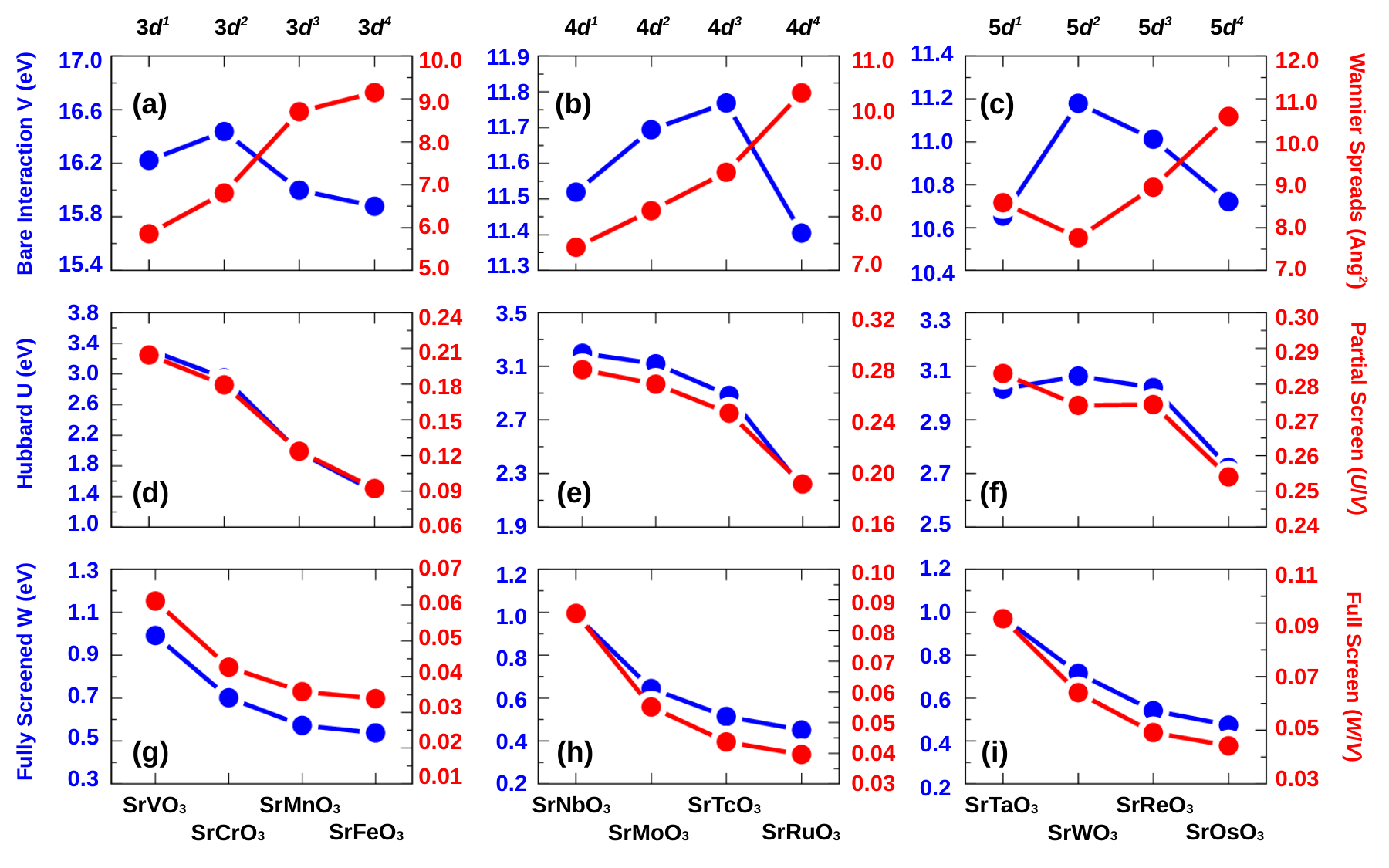}
\caption{Bare interaction $V$ vs. spreads of $t_{2g}$-orbitals (upper panels), Hubbard interaction $U$ (partially screened interaction) vs. partial screening ($U$/$V$) (middle panels), and fully screened interaction $W$ vs. full screening ($W$/$V$) (bottom panels) of Sr$B$O$_3$, within the $t_{2g}$-$t_{2g}$ approximation.}
\label{Fig-SrUJV-t2g}
\end{figure*}

\begin{table}
\begin{ruledtabular}
\caption{DFT calculated lattice parameters that are used for cRPA calculations of cubic perovskites $AM$O$_3$ and energy window $\mathbb{W}$ for $t_{2g}$-$t_{2g}$ and $d$-$dp$ models. Both $t_{2g}$ and $d$-$dp$ Wannier functions are constructed out of the Kohn-Sham states included in $\mathbb{W}$. Because of the strong hybridization between late transition metal e.g.\,Co, Rh and Ir, no distinguishable $t_{2g}$ bands can be projected onto local Wannier functions, the $t_{2g}$-$t_{2g}$ model is not considered for $A$CoO$_3$, $A$RhO$_3$ and $A$IrO$_3$ ($A$=Sr and Ca). Unit of energy window in eV.}
\begin{tabular}{c c c c c}
  & Sr$M$O$_3$  & $a$(\AA) &  $\mathbb{W}_{t_{2g}}$ & $\mathbb{W}_{dp}$ \\
\hline
3$d^1$  & SrVO$_3$   & 3.862 & [-0.98, 1.50] & [-7.18, 5.36]  \\
3$d^2$  & SrCrO$_3$  & 3.823 & [-0.83, 1.53] & [-6.72, 5.36]  \\
3$d^3$  & SrMnO$_3$  & 3.803 & [-1.50, 0.83] & [-7.12, 4.57]  \\
3$d^4$  & SrFeO$_3$  & 3.807 & [-1.60, 0.58] & [-6.89, 4.13]  \\
3$d^5$  & SrCoO$_3$  & 3.815 & -             & [-6.80, 3.60]  \\
\hline\
4$d^1$  & SrNbO$_3$  & 4.055 & [-0.27, 3.56] & [-7.18, 9.03]  \\
4$d^2$  & SrMoO$_3$  & 4.014 & [-0.98, 2.62] & [-7.29, 7.97]  \\
4$d^3$  & SrTcO$_3$  & 3.963 & [-2.02, 1.53] & [-7.79, 7.99]  \\
4$d^4$  & SrRuO$_3$  & 3.953 & [-2.69, 0.77] & [-7.89, 5.81]  \\
4$d^5$  & SrRhO$_3$  & 3.972 & -             & [-7.33, 5.11]  \\
\hline
5$d^1$  & SrTaO$_3$  & 4.062 & [-1.35, 3.03] & [-9.32, 9.13]  \\
5$d^2$  & SrWO$_3$   & 4.013 & [-2.18, 2.50] & [-9.90, 8.14]  \\
5$d^3$  & SrReO$_3$  & 3.978 & [-2.65, 1.42] & [-9.66, 7.37]  \\
5$d^4$  & SrOsO$_3$  & 3.978 & [-2.95, 1.00] & [-9.13, 7.49]  \\
5$d^5$  & SrIrO$_3$  & 3.987 & -             & [-8.73, 6.00]  \\
5$d^6$  & SrPtO$_3$  & 4.036 & -             & [-8.00, 5.21] \\
\hline
\hline
  & Ca$M$O$_3$  & $a$(\AA) &  $\mathbb{W}_{t2g}$ & $\mathbb{W}_{dp}$ \\
\hline
3$d^1$  & CaVO$_3$   & 3.805 & [-1.02, 1.52] & [-7.44, 5.72]  \\
3$d^2$  & CaCrO$_3$  & 3.746 & [-1.54, 0.93] & [-7.69, 5.18]  \\
3$d^3$  & CaMnO$_3$  & 3.724 & [-1.80, 0.54] & [-7.65, 4.68]  \\
3$d^4$  & CaFeO$_3$  & 3.716 & [-1.95, 0.32] & [-7.50, 4.33]  \\
3$d^5$  & CaCoO$_3$  & 3.732 & -             & [-7.14, 4.00]  \\
\hline\
4$d^1$  & CaNbO$_3$  & 4.018 & [-1.40, 2.79] & [-8.32, 8.52]  \\
4$d^2$  & CaMoO$_3$  & 3.954 & [-2.00, 1.79] & [-8.66, 7.45]  \\
4$d^3$  & CaTcO$_3$  & 3.913 & [-2.55, 1.11] & [-8.62, 6.73]  \\
4$d^4$  & CaRuO$_3$  & 3.900 & [-2.84, 0.73] & [-8.24, 6.15]  \\
4$d^5$  & CaRhO$_3$  & 3.917 & -             & [-7.70, 5.46]  \\
\hline
5$d^1$  & CaTaO$_3$  & 4.026 & [-1.45, 3.40] & [-9.43, 9.24]  \\
5$d^2$  & CaWO$_3$   & 3.966 & [-2.41, 2.40] & [-9.90, 8.61]  \\
5$d^3$  & CaReO$_3$  & 3.938 & [-2.75, 1.50] & [-9.83, 7.68]  \\
5$d^4$  & CaOsO$_3$  & 3.931 & [-3.06, 1.00] & [-9.46, 7.00]  \\
5$d^5$  & CaIrO$_3$  & 3.941 & -             & [-9.03, 6.30]  \\
5$d^6$  & CaPtO$_3$  & 3.941 & -             & [-8.27, 5.57] \\
\end{tabular}
\label{table1}
\end{ruledtabular}
\end{table}

\begin{table*}
\begin{ruledtabular}
\caption{Bare $V$, Hubbard $U$ and fully screened $W$ interactions, and corresponding Hund's exchange $J_{bare}$, $J$ and $J_{screened}$ between $d$ orbitals within $d$-$dp$ approximation for Sr$M$O$_3$ ($M$=V, Cr, Mn, Fe, Co, Nb, Mo, Tc, Ru, Rh, Ta, W, Re, Os, Ir, Pt) perovskites. Their Wannier orbital spreads are also shown. All the energy units are in eV.}
\begin{tabular}{c | c c c c c c | c c c c c c}
  & $d^1$ & $d^2$  & $d^3$ &  $d^4$ &  $d^5$ & $d^6$  & $d^1$ & $d^2$  & $d^3$ & $d^4$ & $d^5$ & $d^6$ \\
\hline
3$d$ & SrVO$_3$ & SrCrO$_3$ & SrMnO$_3$ & SrFeO$_3$ & SrCoO$_3$ & &  CaVO$_3$ & CaCrO$_3$ & CaMnO$_3$ & CaFeO$_3$ & CaCoO$_3$ & \\
\hline
 $U$            & 3.24 & 2.64 & 2.79 & 3.48 & 2.70 &       & 3.42 & 2.85 & 2.85 & 4.88 & 2.84  \\
 $J$            & 0.62 & 0.63 & 0.66 & 0.69 & 0.73 &       & 0.64 & 0.65 & 0.67 & 0.70 & 0.74  \\
\hline
 $V$            & 20.88& 22.25& 23.62& 24.91& 26.19&       & 21.09& 22.54& 23.67& 25.00& 26.28 \\
 $J_{bare}$     & 0.74 & 0.76 & 0.80 & 0.84 & 0.88 &       & 0.75 & 0.77 & 0.80 & 0.84 & 0.88  \\
\hline
 $W$            & 1.51 & 1.22 & 1.15 & 1.17 & 1.29 &       & 1.57 & 1.27 & 1.18 & 1.19 & 1.32  \\
 $J_{screened}$ & 0.55 & 0.51 & 0.50 & 0.53 & 0.58 &       & 0.56 & 0.52 & 0.51 & 0.53 & 0.59  \\
\hline
 $U'$           & 1.97 & 1.36 & 1.44 & 2.05 & 1.22 &       & 2.12 & 1.54 & 1.50 & 3.44 & 1.34  \\
\hline
 $U/V$          &15.5\%&11.8\%&11.8\%&13.9\%&10.3\%&       &16.2\%&12.6\%&12.0\%&19.5\%&10.8\% \\
 $W/V$          &7.2\% &5.5\% &4.8\% &4.6\% &4.9\% &       &7.4\% &5.6\% &4.9\% &4.7\% &5.0\%  \\
\hline
 Wannier Spreads& 3.34 & 3.06 & 2.69 & 2.41 & 2.19 &       & 3.10 & 2.75 & 2.72 & 2.40 & 2.17 \\
\hline
\hline
4$d$  & SrNbO$_3$ & SrMoO$_3$ & SrTcO$_3$ & SrRuO$_3$ & SrRhO$_3$ & & CaNbO$_3$ & CaMoO$_3$ & CaTcO$_3$ & CaRuO$_3$ & CaRhO$_3$ \\
\hline
 $U$            & 3.54 & 3.09 & 3.03 & 3.40 & 2.86 &       & 3.22 & 2.96 & 2.97 & 3.46 & 2.83 \\
 $J$            & 0.46 & 0.49 & 0.52 & 0.53 & 0.54 &       & 0.47 & 0.50 & 0.52 & 0.54 & 0.54 \\
\hline
 $V$            & 14.75& 15.57& 16.54& 17.29& 18.12&       & 14.80& 15.73& 16.52& 17.42& 18.26 \\
 $J_{bare}$     & 0.58 & 0.61 & 0.64 & 0.67 & 0.70 &       & 0.58 & 0.61 & 0.64 & 0.67 & 0.71 \\
\hline
 $W$            & 1.59 & 1.20 & 1.13 & 1.16 & 1.59 &       & 1.63 & 1.25 & 1.14 & 1.18 & 1.41  \\
 $J_{screened}$ & 0.42 & 0.41 & 0.40 & 0.40 & 0.43 &       & 0.43 & 0.42 & 0.41 & 0.41 & 0.43  \\
\hline
 $U'$           & 2.58 & 2.08 & 1.97 & 2.29 & 1.71 &       & 2.27 & 1.94 & 1.92 & 2.34 & 1.71  \\
\hline
 $U/V$          &24.0\%&19.8\%&18.3\%&19.6\%&15.7\%&       &21.7\%&18.8\%&18.0\%&19.8\%&15.5\%  \\
 $W/V$          &10.7\%&7.7\% &6.8\% &6.7\% &8.7\% &       &11.0\%&7.9\% &6.9\% &6.7\% &7.7\%   \\
\hline
Wannier Spreads & 5.78 & 5.35 & 4.66 & 4.42 & 4.12 &       & 5.73 & 5.05 & 4.77 & 4.21 & 3.83 \\
\hline
\hline
5$d$ & SrTaO$_3$ & SrWO$_3$ & SrReO$_3$ & SrOsO$_3$ & SrIrO$_3$ & SrPtO$_3$ & CaTaO$_3$ & CaWO$_3$ & CaReO$_3$ & CaOsO$_3$ & CaIrO$_3$ & CaPtO$_3$\\
\hline
 $U$            & 3.48 & 3.17 & 3.12 & 3.29 & 3.36 & 2.34 & 2.94 & 2.86 & 2.96 & 3.17 & 3.24 & 2.47 \\
 $J$            & 0.45 & 0.47 & 0.48 & 0.50 & 0.52 & 0.50 & 0.45 & 0.47 & 0.48 & 0.50 & 0.52 & 0.51 \\
\hline
 $V$            & 13.71& 14.44& 14.94& 15.63& 16.13& 16.69& 13.81& 14.52& 15.08& 15.67& 16.26& 16.78 \\
 $J_{bare}$     & 0.55 & 0.58 & 0.60 & 0.63 & 0.65 & 0.67 & 0.56 & 0.59 & 0.61 & 0.63 & 0.66 & 0.68\\
\hline
 $W$            & 1.60 & 1.28 & 1.13 & 1.16 & 1.30 & 1.62 & 1.66 & 1.32 & 1.16 & 1.16 & 1.35 & 1.71\\
 $J_{screened}$ & 0.41 & 0.40 & 0.39 & 0.39 & 0.40 & 0.44 & 0.42 & 0.41 & 0.40 & 0.40 & 0.40 & 0.44\\
\hline
 $U'$           & 2.55 & 2.20 & 2.12 & 2.24 & 2.28 & 1.28 & 2.02 & 1.89 & 1.95 & 2.12 & 2.16 & 1.39\\
\hline
 $U/V$          &25.3\%&21.9\%&20.9\%&21.0\%&20.8\%&14.0\%&21.3\%&19.6\%&19.6\%&20.2\%&19.9\%&14.7\% \\
 $W/V$          &11.6\%&8.8\% &7.5\% &7.4\% &8.0\% &9.7\% &12.0\%&9.0\% &7.7\% &7.4\% &8.3\% &10.1\% \\
\hline
Wannier Spreads & 6.99 & 6.14 & 6.08 & 5.35 & 5.20 & 4.93 & 6.74 & 5.96 & 5.65 & 5.26 & 4.93 & 4.75\\
\end{tabular}
\label{table2}
\end{ruledtabular}
\end{table*}

\begin{table*}
\begin{ruledtabular}
\caption{Bare $V$, Hubbard $U$ and fully screened $W$ interactions, and corresponding Hund's exchange $J_{bare}$, $J$ and $J_{screened}$ between $t_{2g}$ orbitals within $t_{2g}$-$t_{2g}$ approximation for Sr$M$O$_3$ ($M$=V, Cr, Mn, Fe, Co, Nb, Mo, Tc, Ru, Rh, Ta, W, Re, Os, Ir, Pt) perovskites. Their Wannier orbital spreads are also shown. All the energy units are in eV.}
\begin{tabular}{c | c c c c | c c c c }
  & $d^1$ & $d^2$  & $d^3$ &  $d^4$ & $d^1$ & $d^2$  & $d^3$ & $d^4$ \\
\hline
3$d$ & SrVO$_3$ & SrCrO$_3$ & SrMnO$_3$ & SrFeO$_3$ &  CaVO$_3$ & CaCrO$_3$ & CaMnO$_3$ & CaFeO$_3$  \\
\hline
 $U$            & 3.32 & 2.95 & 1.98 & 1.46 & 3.38 & 3.08 & 2.24 & 1.70   \\
 $J$            & 0.46 & 0.43 & 0.38 & 0.36 & 0.46 & 0.44 & 0.40 & 0.38    \\
\hline
 $V$            & 16.22 & 16.43 & 16.00 & 15.88 & 16.20 & 16.54 & 16.18 & 16.19    \\
 $J_{bare}$     & 0.56 & 0.54 & 0.50 & 0.48 & 0.56 & 0.54 & 0.51 & 0.49    \\
\hline
 $W$            & 0.99 & 0.71 & 0.57 & 0.54 & 1.01 & 0.72 & 0.59 & 0.56    \\
 $J_{screened}$ & 0.36 & 0.30 & 0.26 & 0.26 & 0.36 & 0.31 & 0.27 & 0.27    \\
\hline
 $U'$           & 2.34 & 2.00 & 1.10 & 0.63 & 2.41 & 2.13 & 1.33 & 0.82    \\
\hline
 $U/V$          &20.4\%&17.9\%&12.3\%&9.2\% &20.8\%&18.6\%&13.8\%&10.4\%    \\
 $W/V$          &6.1\% &4.2\% &3.5\% &3.3\% &6.2\% &4.3\% &3.6\% &3.4\%     \\
\hline
Wannier Spreads & 5.85 & 6.80 & 8.70 & 9.15 & 6.00 & 6.733 & 8.50 & 9.09    \\
\hline
\hline
4$d$  & SrNbO$_3$ & SrMoO$_3$ & SrTcO$_3$ & SrRuO$_3$ & CaNbO$_3$ & CaMoO$_3$ & CaTcO$_3$ & CaRuO$_3$  \\
\hline
 $U$            & 3.20 & 3.12 & 2.88 & 2.19 & 2.80 & 2.92 & 2.87 & 2.54    \\
 $J$            & 0.33 & 0.33 & 0.31 & 0.28 & 0.29 & 0.31 & 0.31 & 0.29    \\
\hline
 $V$            & 11.52 & 11.69 & 11.77 & 11.40 & 10.41 & 11.35 & 11.74 & 11.53    \\
 $J_{bare}$     & 0.42 & 0.41 & 0.39 & 0.35 & 0.36 & 0.39 & 0.39 & 0.36    \\
\hline
 $W$            & 0.99 & 0.64 & 0.51 & 0.45 & 0.88 & 0.64 & 0.52 & 0.46    \\
 $J_{screened}$ & 0.27 & 0.22 & 0.19 & 0.17 & 0.23 & 0.22 & 0.19 & 0.17    \\
\hline
 $U'$           & 2.52 & 2.43 & 2.20 & 1.54 & 2.23 & 2.26 & 2.19 & 1.87    \\
\hline
 $U/V$          &27.7\%&26.6\%&24.4\%&19.2\%&26.9\%&25.7\%&24.4\%&21.9\%    \\
 $W/V$          &8.5\% &5.5\% &4.3\% &3.9\% &8.4\% &5.6\% &4.4\% &4.0\%     \\
\hline
Wannier Spreads & 7.43 & 8.11 & 8.83 & 10.31 & 11.01 & 9.21 & 9.00 & 10.40    \\
\hline
\hline
5$d$ & SrTaO$_3$ & SrWO$_3$ & SrReO$_3$ & SrOsO$_3$ & CaTaO$_3$ & CaWO$_3$ & CaReO$_3$ & CaOsO$_3$\\
\hline
 $U$            & 3.01 & 3.06 & 3.02 & 2.72 & 2.39 & 1.97 & 3.44 & 3.26    \\
 $J$            & 0.32 & 0.32 & 0.30 & 0.27 & 0.34 & 0.30 & 0.29 & 0.28    \\
\hline
 $V$            & 10.65 & 11.18 & 11.01 & 10.72 & 11.23 & 10.68 & 10.79 & 10.74    \\
 $J_{bare}$     & 0.40 & 0.40 & 0.38 & 0.35 & 0.42 & 0.38 & 0.37 & 0.35    \\
\hline
 $W$            & 0.98 & 0.71 & 0.54 & 0.47 & 0.95 & 0.65 & 0.55 & 0.48    \\
 $J_{screened}$ & 0.26 & 0.23 & 0.19 & 0.17 & 0.27 & 0.22 & 0.19 & 0.17    \\
\hline
 $U'$           & 2.39 & 2.38 & 2.36 & 2.09 & 1.74 & 1.39 & 2.78 & 2.60    \\
\hline
 $U/V$          &28.2\%&27.4\%&27.4\%&25.4\%&21.3\%&18.4\%&31.8\%&30.3\%   \\
 $W/V$          &9.1\% &6.4\% &4.9\% &4.4\% &8.4\% &6.0\% &5.0\% &4.5\%    \\
\hline
Wannier Spreads & 8.58 & 7.76 & 8.94 & 10.59 & 6.92 & 9.52 & 9.74 & 10.60    \\
\end{tabular}
\label{table3}
\end{ruledtabular}
\end{table*}

The DFT calculated non-magnetic band structures of Sr-based 3$d$-5$d$ $AB$O$_3$ are displayed in Figs.~\ref{Fig-Srband1}-\ref{Fig-Srband3} (see Fig.~\ref{Fig-Caband1}-\ref{Fig-Caband3} in Appendix for the bands of Ca-based $AB$O$_3$), respectively. The DFT optimized lattice parameters are listed in Table~\ref{table1}. 
One can see from Fig.~\ref{Fig-Srband1} that as the $d$-band filling increases from 3$d^1$ to 3$d^5$, several distinct trends become apparent: (1) The $t_{2g}$ bandwidths shrink, and their energies shift downward, coming into contact with O-2$p$ bands from 3$d^3$ (SrMnO$_3$) and 3$d^4$ (SrFeO$_3$). (2) The $e_g$ bands also shrink and shift to lower energies, making contact with $t_{2g}$ bands starting from 3$d^3$ (SrMnO$_3$). These factors lead to a significant reduction in the total $d$-bandwidth with increasing $d$-band filling from $d^1$-$d^5$. (3) The O-2$p$ bands, spanning from approximately -7.0\,eV to -2.0\,eV for 3$d^1$ SrVO$_3$ and -7.0\,eV to 0.0\,eV for 3$d^5$ SrCoO$_3$, shift upward towards the Fermi energy with band filling, eventually overlapping with $d$-bands. (4) The Sr-$d$ bands, initially above $\sim$4\,eV, shift upward with increasing $d$-band filling. (5) For 3$d^1$ SrVO$_3$ to 3$d^3$ SrMnO$_3$, distinct and separated $t_{2g}$ bands are obtained; however, for both 3$d^4$ SrFeO$_3$ and 3$d^5$ SrCoO$_3$, their $t_{2g}$ bands undergo strong hybridization with O-2$p$ bands, making the projection of DFT bands onto $t_{2g}$-based Wannier orbitals impractical.

As a consequence, when the $B$-site atomic number increases, the $d$-orbitals undergo bandwidth reduction while the distance between between $d$- and $p$-band decreases, leading to an enhancement in hybridization between $d$- and $p$-orbitals from early to late $AB$O$_3$. This trend is further supported by the decrease in the $d$-$p$ charge transfer energy ($\Delta_{dp}$), in qualitative agreement with optics experiments \cite{PhysRevB.67.113101}. Additionally, the charge transfer energy is larger in 5$d$ and 4$d$ than in 3$d$, which is attributed to the larger orbital extension of the former (compare Fig.~\ref{Fig-Srband3}and Fig.~\ref{Fig-Srband2} to Fig.~\ref{Fig-Srband1}).

\subsection{B. $d$-$dp$ approximation for Sr-based 3$d$ series}

The cRPA derived interaction parameters for the 3$d$-5$d$ Sr-based $AB$O$_3$ series  using the $d$-$dp$ model are presented in Table~\ref{table2} and Fig.~\ref{Fig-SrUJV-dp}.

We start from analyzing the tendencies of the interactions evolution with the atomic number in Sr-based 3$d$ series. In Fig.~\ref{Fig-SrUJV-dp}, the $d$-$dp$ interactions for the bare ($V$), partially ($U$), and fully screened ($W$) Coulomb interactions of Sr$B$O$_3$ are shown. As band filling and atomic number increase within the same period of the periodic table (3$d$), the bare interactions $V$ monotonously increases from 20.9\,eV (3$d^1$ SrVO$_3$) to 26.2\,eV (3$d^5$ SrCoO$_3$) [Fig.~\ref{Fig-SrUJV-dp}(a)]. 
The values obtained using the $d$-$dp$ model are larger than the previous estimates based on the $t_{2g}$-$t_{2g}$ approximation, such as 15.8\,eV \cite{PhysRevB.99.155143} and 16.1\,eV \cite{PhysRevB.86.165105} for SrVO$_3$, but are closer to the value derived from the $d$-$dp$ model, which is, for instance, 19.5\,eV for SrVO$_3$ using the (L)APW+lo framework in Ref.~\cite{PhysRevB.86.165105}.
Additionally, the degree of orbital localization of $d$-orbitals ($t_{2g}$+$e_g$ basis), which varies with the inverse of the Wannier $d$-orbital spreads, monotonously increases. Thus, the less extended the orbitals, the higher the bare Coulomb repulsion $V$. This means that the values of bare interactions are merely decided by the localization of correlated orbitals (i.e., the $d$ orbitals for the $d$-$dp$ model). Besides the Wannier orbital spreads, another measure of orbital localization is the bandwidth (Fig.~\ref{Fig-Srband1}, bottom panels). From 3$d^1$ to 3$d^5$, the $d$-bands progressively approach the O-2$p$ bands, eventually overlapping at 3$d^4$ SrFeO$_3$, which results in increased $d$-$p$ hybridization. This leads to a noticeable reduction in $d$-bandwidth, evident in both the band structures and density of states (DOS) plots (as shown in Fig.~\ref{Fig-Srband1}-\ref{Fig-Srband3}).

Next, let us delve into the evolution of the fully screened interaction $W$. Unlike the bare interaction $V$, $W$ accounts for electronic screening from all screening channels resulting from electronic polarization. Larger screening effects lead to stronger reduction from $V$ to $W$. 
The values we obtained are highly consistent with previous reports for 3$d$ Sr$B$O$_3$ systems, such as SrVO$_3$, SrCrO$_3$ and SrMnO$_3$ \cite{PhysRevB.86.165105}, for which the $W$ are in the region between 1.0-1.5\,eV.
Interestingly, $W$ for the Sr-based 3$d$ series initially decreases from 1.5\,eV (3$d^1$ SrVO$_3$) to 1.2\,eV (3$d^3$ SrMnO$_3$), 
then increases from 3$d^3$ to 3$d^5$ SrCoO$_3$ (1.3\,eV), reaching a minimum at 3$d^3$ SrMnO$_3$. This indicates a consequence of a significantly increased full screening from 3$d^1$ to 3$d^3$, counteracting the enhancement of orbital localization and bare $V$. However, from 3$d^3$ to 3$d^5$, the screening effect remains relatively constant.

The local minimum of $W$ at 3$d^3$ indicates a predominant influence of increasing screening effect from 3$d^1$ to 3$d^3$. This observation aligns with the reduction in bandwidth within the 3$d$ series. In correlated systems, full electronic screening (mediating $V$ to $W$) involves the creation of particle-hole and plasma excitation. At the RPA level,  the strength of full screening is inversely proportional to the energy difference between occupied and unoccupied states.
Examining the DFT bands in Fig.~\ref{Fig-Srband1}(a-e), we observe a reduction in bandwidth as $d$-electron number increases, causing the empty and occupied states to approach each other near the Fermi energy. The diminished separation between occupied and empty states leads to a stronger full screening effect, and consequently, a smaller $W$. This bandwidth reduction explains the decrease in $W$ and the enhancement of full screening from 3$d^1$ to 3$d^3$. The increasing screening, in turn, counteracts the tendency towards more localized 3$d$-orbitals and the enhancement of bare interaction $V$.

From 3$d^3$ (SrMnO$_3$) to 3$d^5$ (SrCoO$_3$), $W$ increases from 1.15\,eV to 1.29\,eV. This can be understood from two aspects: (1) the bare interaction $V$ already increases to a large value of 26.19\,eV in 3$d^5$, and even though the full screening in the localized 3$d$ series is significant, it fails to counteract the effect of increasing localization; (2) as the atomic number and 3$d$-band filling increase, the unoccupied states gradually become occupied due to the shift of the Fermi energy. This decreases the possibility of creating particle-hole excitations around E$_f$, counteracting the enhancement of the full screening effect.
To quantify the full screening strength in 3$d$ series, we calculated the ratio between $W$ and $V$, as shown in Fig.~\ref{Fig-SrUJV-dp}(g) and Table~\ref{table2}. We obtained values of 0.072, 0.055, 0.048, 0.046, 0.049 from 3$d^1$ SrVO$_3$ to 3$d^5$ SrCoO$_3$, respectively. The tendency of $W$/$V$ is basically consistent with that of $W$, indicating that $W$ is mediated by full screening strength.

Before inspecting the trend of the partially screened Coulomb interaction $U$, we first classify the full screening effect in the Sr-based $AB$O$_3$ series into three major contributions: (1) the $d$-$d$ screening, i.e., the screening from occupied $d$-$t_{2g}$ states to unoccupied $t_{2g}$ and $e_g$ states; (2) the $d$-$p$ screening, i.e., the screening from occupied O-2$p$ states to unoccupied $d$ (mostly $e_g$ in $AB$O$_3$) states; (3) the rest screening, which (mostly) consists of the screening from occupied $d$ and O-2$p$ states to the higher unoccupied Sr-4$d$ states. After removing the $d$-$d$ contribution from the full screening [which is the summation of (1-3)], the remaining partial screening consists of the (2) $d$-$p$ and (3) rest screening. Unlike $V$ and $W$, the partially screened interaction, i.e., the Hubbard $U$, exhibits an unusual tendency: two maximum values of $U$ are found at 3$d^1$ SrVO$_3$ and 3$d^4$ SrFeO$_3$, respectively. To quantify the strength of partial screening, we also calculate the ratio between $U$ and $V$, as shown in Fig.~\ref{Fig-SrUJV-dp}(d) and Table~\ref{table2}.

From 3$d^1$ to 3$d^2$, $U$ decreases from 3.25\,eV to 2.64\,eV. This reduction agrees with the tendency of $W$, indicating that $d$-$p$ screening dominates the process from 3$d^1$ to 3$d^2$.
A similar reduction in $U$ from SrVO$_3$ to SrCrO$_3$, based on the $t_{2g}$+$p$ model, was previously observed in Ref.~\cite{PhysRevB.86.165105} and recently reported in Ref.~\cite{mushkaev2024internal}.
Please note that for both SrVO$_3$ and SrCrO$_3$, the $e_g$ bands are well-separated from the $t_{2g}$ bands [see Fig.~\ref{Fig-Srband1}(a,b)]. Therefore, including $e_g$ bands in the model for cRPA calculations is not expected to significantly influence the resulting values.
From 3$d^2$ to 3$d^3$, the $U$ remains almost constant with a slight enhancement from 2.64\,eV to 2.79\,eV. The partial screening strength ($U$/$V$) for 3$d^2$ and 3$d^3$ is the same, indicated by $U$/$V$ as 0.118 (Table~\ref{table2}). This means that the summation of rest screening and $d$-$p$ screening in 3$d^2$ SrCrO$_3$ and 3$d^3$ SrMnO$_3$ are comparable. However, the full screening strength in 3$d^3$ is obviously stronger than that in 3$d^2$ SrCrO3 [Fig.~\ref{Fig-SrUJV-dp}(g) and Table~\ref{table2}: $W$/$U$ is 0.055 for 3$d^2$ and 0.048 for 3$d^3$]. Hence, we conclude that the $d$-$d$ screening, excluded in the calculations for Hubbard $U$, is much stronger in 3$d^3$ SrMnO$_3$ than in 3$d^2$ SrCrO$_3$. This can be explained by their electronic structures [Fig.~\ref{Fig-Srband1}(b-c)]: in the bands of SrCrO$_3$ and SrMnO$_3$, the $d$-bandwidth of 3$d^3$ SrMnO$_3$ is obviously reduced compared with that of 3$d^2$ SrCrO$_3$. This shortens the energetic distance between occupied and unoccupied $d$-states, leading to an enhanced $d$-$d$ screening. Moreover, as shown in the DOSs (bottom panels of Fig.~\ref{Fig-Srband1}): when one more electron is distributed to the $t_{2g}$ orbital in 3$d^3$ SrMnO$_3$, more $t_{2g}$ states are occupied, increasing the possibility of forming particle-hole excitations in $d$-orbitals. This explains the stronger $d$-$d$ screening in 3$d^3$ than in 3$d^2$ and why there are comparable summations of $d$-$p$ and rest screening in both 3$d^2$ and 3$d^3$. Considering that the bare $V$ is 23.62\,eV in 3$d^3$ SrMnO3 and 22.26\,eV in 3$d^2$ SrCrO$_3$, the Hubbard $U$ in SrMnO$_3$ is slightly larger (0.149\,eV) than in SrCrO$_3$.

The transition from 3$d^3$ SrMnO$_3$ to 3$d^4$ SrFeO$_3$ exhibits an unusual behavior. The Hubbard $U$ in 3$d^4$ is 3.48\,eV, approximately 0.69\,eV larger than that of 3$d^3$ SrMnO$_3$ (2.79\,eV). This local maximum of $U$ indicates that the summation of $d$-$p$ and rest screening is significantly weaker in 3$d^4$ than in 3$d^3$. Moreover, the full screening strength ($W$/$V$) is stronger in 3$d^4$ than in 3$d^3$ (as shown in Table~\ref{table2}: $W$/$U$ is 0.046 for 3$d^4$ and 0.048 for 3$d^3$). Hence, the $d$-$d$ screening in 3$d^4$ represents a local maximum among the 3$d$ series. This is explained by the DOSs in Fig.~\ref{Fig-Srband1}. For $3d^4$, both the $t_{2g}$ and $e_g$ states overlap around $E_f$, and the $t_{2g}$ states are almost fully filled, the DOS peak located at $E_f$ (similar to 4$d^4$ SrRuO$_3$). The number of occupied and unoccupied states is comparable (4 electrons and 6 holes in $d$), resulting in a larger possibility of particle-hole excitation and $d$-$d$ screening.

From 3$d^4$ to 3$d^5$ SrCoO$_3$, the Hubbard $U$ drops to 2.70\,eV, indicating that 3$d^5$ exhibits the strongest partial screening strength ($d$-$p$ plus the rest screening) among 3$d^1$ to 3$d^5$ and the corresponding weakest $d$-$d$ screening. In 3$d^5$, the $d$-$p$ screening dominates the $U$, and the exclusion of $d$-$d$ screening in $d$-$dp$ approximation has a tiny influence on the physics and does not significantly differ $U$ and $W$. This can be explained by the DOSs in Fig.~\ref{Fig-Srband1}(e). The O-2$p$ states form three peaks in the entire energy region. For instance, in Fig.~\ref{Fig-Srband1}(a) showing the DOS of 3$d^1$ SrVO$_3$, three O-2$p$ peaks are found at approximately -5.5\,eV, -3.5\,eV, and 0.5\,eV, respectively. The two lower O-2$p$ peaks are formed by the O-2$p$ bands, while the one near $E_f$ is the hybridization peak with obvious $V$-3$d$ orbital characters. As the $d$-electron increases, the hybridization peak gradually shifts down, while the O-2$p$ peaks below $E_f$ shift up. Approaching 3$d^5$ SrCoO$_3$ [Fig.~\ref{Fig-Srband1}(e)], the upper O-2$p$ peak approaches the hybridization peak and finally overlaps with it in $3d5$ SrCoO$_3$, forming a higher O-2$p$ peak near $E_f$. The $d$-$p$ screening contribution is significantly enhanced by the formation of this peak, becoming the largest contribution instead of the $d$-$d$ screening to the full screening. Even when $d$-$d$ screening is excluded in the $d$-$dp$ model, the remaining $d$-$p$ screening still plays an effective role in reducing the $V$ to a smaller $U$.

\subsection{C. $d$-$dp$ approximation for Sr-based 4$d$ and 5$d$ series}

As shown in Fig.~\ref{Fig-SrUJV-dp}(b,e,h) and Table~\ref{table2}, the trend of $V$, $W$, and $U$ in the Sr-based 4$d$ $AB$O$_3$ are roughly similar to those in the 3$d$ ones. The bare $V$ in 4$d$ are generally smaller than those in 3$d$, given the more extended nature of 4$d$ orbitals, and our previous discussion confirms that the bare $V$ is primarily determined by the degree of $d$-orbital localization.

The behaviors of Hubbard $U$ in 4$d$ $AB$O$_3$ are akin to those in 3$d$ ones, with two maxima observed at $4d^1$ SrNbO$_3$ and 4$d^4$ SrRuO$_3$. The $U$ decreases initially from 4$d^1$ (3.55\,eV) to 4$d^3$ SrTcO$_3$ (3.03\,eV), then increases to 3.40\,eV at 4$d^4$ SrRuO$_3$, and finally decreases to 2.86\,eV at 4$d^5$ SrRhO$_3$. 
The resulting $U$ values for the 4$d$ series, for instance 4$d^1$ SrNbO$_3$, is consistent with the 3.2\,eV reported in Ref.~\cite{selisko2024dynamical} and 3.0\,eV reported in Ref.~\cite{PhysRevB.86.165105}, and for SrMoO$_3$ (with $t_{2g}$-$t_{2g}$ approximation), 2.8\,eV \cite{PhysRevMaterials.1.043803} and 3.12\,eV (with $d$-$dp$ approximation) in Ref.~\cite{PhysRevB.104.035102}.
The increase of $U$ from 4$d^3$ to 4$d^4$ is only 0.37\,eV smaller than the increase from 3$d^3$ to 3$d^4$ (0.69 eV). The stronger crystal field and delocalization of 4$d$ induce a larger splitting between $t_{2g}$ and $e_g$ states compared with those in 3$d$, resulting in weaker screening effect from unoccupied states. Therefore, the $d$-$d$ contribution in 4$d$ is smaller than in 3$d$, and excluding the $d$-$d$ screening induces a less pronounced enhancement of $U$ from 4$d^3$ to 4$d^4$.

Some notable observations include: (1) The fully screened interaction $W$ in 4$d$ are comparable to those of 3$d$ $AB$O$_3$, but the (full) screening effect in 3$d$ is significantly stronger than in 4$d$; (2) Despite the generally smaller bare $V$ in 4$d$ than in 3$d$ due to less localized $d$-orbitals, the $U$ in 4$d$ is approximately 0.2\,eV larger than in 3$d$, indicating that the summations of $d$-$p$ and rest screening in 4$d$ are remarkably weaker than in 3$d$ $AB$O$_3$; (3) The only difference in the trends of Hubbard $U$ between 3$d$ and 4$d$ series occurs from $d^2$ to $d^3$: in 3$d$ $AB$O$_3$, 3$d^2$ SrCrO$_3$ and 3$d^3$ SrMnO$_3$ host roughly same partial screen strength ($U$/$V$), while in 4$d$, the $U$/$V$ in 4$d^2$ SrMoO$_3$ is weaker than in 4$d^3$ SrTcO$_3$, hinting that the enhanced $d$-$d$ screening from 4$d^2$ to 4$d^3$ is not as pronounced as in the 3$d$ series due to the larger crystal field splitting between $t_{2g}$ and $e_g$ states.

In 5$d$ $AB$O$_3$, the bare interaction $V$ is even smaller compared to 4$d$ and 3$d$ series due to the less localized nature of 5$d$ orbitals. Additionally, both bare $V$ and fully screened $W$ exhibit similar behaviors as in 3$d$ and 4$d$ series. However, the trend of Hubbard $U$ is slightly different. The maximum value of partial screen strength $U$/$V$ is still located at 5$d^4$ SrOsO$_3$, but the amplitude of the enhancement from 5$d^3$ to 5$d^4$ is smaller compared with the jumps in 3$d$ and 4$d$. The configurations from 5$d^3$ to 5$d^5$ host roughly the same strength of partial screenings, as shown in Fig.~\ref{Fig-SrUJV-dp}(f) and Table~\ref{table2}. The bare $V$ increases from 14.94\,eV (5$d^3$ SrReO$_3$) to 15.64\,eV (5$d^4$ SrOsO$_3$) and 16.14\,eV (5$d^5$ SrIrO$_3$). Hence, the $U$ slightly increases from 3.13\,eV to 3.37\,eV from 5$d^3$ to 5$d^5$. At 5$d^6$ SrPtO$_3$, $U$ drops to 2.34\,eV because the $d$-$p$ screening contributes more than $d$-$d$ screening, as the O-2$p$ peak below $E_f$ merges with the hybridization peak at $E_f$, forming a high peak and enhancing the $d$-$p$ screening.

{\em Hund's interaction}.
The bare ($J_{ bare}$), partially screened ($J$), and fully screened ($J_{screened}$) Hund's exchanges are derived from the off-diagonal matrix elements of the interaction matrix (Table~\ref{table2}). The evolution of Hund's interactions for 3$d$, 4$d$, and 5$d$ series follows the same tendencies, and we will use 3$d$ as an example.
In the 3$d$ Sr$B$O$_3$ series, partially screened Hund's exchange $J$ monotonously increases from 0.63\,eV (3$d^1$ SrVO$_3$) to 0.73\,eV (3$d^5$ SrCoO$_3$), exceeding the $J$ within the $t_{2g}$-$t_{2g}$ approximation (Table~\ref{table3}). These values align with the $J$ values commonly used in previous DFT+$U$ or DFT+DMFT calculations for $AB$O$_3$ and $AB$O$_2$ materials \cite{PhysRevLett.93.156402,PhysRevB.85.205133,PhysRevB.93.235109,PhysRevLett.124.166402}. Similar to partially screened $J$, the bare exchange interaction $J_{ bare}$ also monotonously increases with atomic number from 0.74\,eV (3$d^1$ SrVO$_3$) to 0.88\,eV (3$d^5$ SrCoO$_3$). This indicates that both $J$ and $J_{ bare}$ are determined by the degree of $d$-orbital localization, rather than following the tendencies of $W$ and $U$. The more localized the $d$-orbitals, the stronger $J$ and $J_{bare}$. In contrast to $J$ and $J_{bare}$, the fully screened $J_{screened}$ follows the behavior of Coulomb interactions, i.e., the fully screened $W$, with a local minimum at 3$d^3$ SrMnO$_3$. The $J_{screened}$ values are 0.55\,eV for 3$d^1$ SrVO$_3$, 0.51\,eV for 3$d^2$ SrCrO$_3$, 0.50\,eV for 3$d^3$ SrMnO$_3$, 0.53\,eV for 3$d^4$ SrFeO$_3$, and 0.59\,eV for 3$d^5$.

The above discussion about $J$ implies that only the full screening effect significantly modifies the $J$ values, while both $J$ and $J_{ bare}$ are primarily influenced by orbital localization. It is worth noting that, compared with $U$, $W$, and $V$ ($\sim$1.0\,eV), the amplitudes of $J_{ bare}$, $J$, and $J_{screened}$ are an order of magnitude smaller ($\sim$0.1\,eV). Therefore, quantitative comparisons between the same period may not be as instructive as  Coulomb interactions.

Finally, we delve into the outcomes of intra-orbital Coulomb repulsion $U'$, as presented in Table~\ref{table2}. For 3$d$, 4$d$, and 5$d$ series, $U'$ consistently yields $U'$=$U$-2$J$ due to the rotational symmetry, confirming the robustness of our cRPA calculations. The behavior of $U'$ mirrors that of $U$.

\subsection{D. $t_{2g}$-$t_{2g}$ approximation for Sr-based 3$d$ serie}

Within the $t_{2g}$-$t_{2g}$ approximation, the $t_{2g}$-projected local orbitals within the energy window $\mathbb{W}_{t_{2g}}$ (Table~\ref{table1}) result in ``more extended'' $t_{2g}$ Wannier orbitals. The charge transfer energy and the hybridization between the $t_{2g}$ and O-2$p$ bands contribute to the finite weight/tail of the $t_{2g}$ Wannier functions at the O sites. Consequently, a smaller $d$-$p$ charge transfer energy ($\Delta
_{dp}$) leads to more extended projected $t_{2g}$ orbitals. This tendency is evident in the 3$d$ Sr$B$O$_3$ bands and $t_{2g}$ Wannier orbital spreads: as the $\Delta
_{dp}$ decreases from 3$d^1$ SrVO$_3$ to 3$d^4$ SrFeO$_3$ (as illustrated in Fig.~\ref{Fig-Srband1} by the splitting between $d$ and O-2$p$ states), the $t_{2g}$ orbitals become more extended accordingly [as depicted in Fig.~\ref{Fig-SrUJV-t2g}(a) with larger Wannier spreads].

The interaction parameters obtained from $t_{2g}$-$t_{2g}$ approximation are detailed in Table~\ref{table3}, and their evolution and the relationship between the most effective factor within the series are illustrated in Fig.~\ref{Fig-SrUJV-t2g}.
In Fig.~\ref{Fig-SrUJV-t2g}(a-c) and Table~\ref{table3}, the unscreened bare interaction $V_{t_{2g}}$ within the $t_{2g}$-$t_{2g}$ model does not increase with $d$-electron count as $V_{d-dp}$ does in $d$-$dp$ model.
Additionally, the values of $V_{t_{2g}}$ are generally smaller than the values of $V_{d-dp}$ due to the greater delocalization of projected $t_{2g}$ orbitals in the $t_{2g}$-$t_{2g}$ model.
For more comparison between ours and previous computations on the interaction parameters for the $AB$O$_3$ compounds based on the $t_{2g}$-$t_{2g}$ approximation, we refer the readers to the references in Refs.~\cite{PhysRevB.86.165105,PhysRevB.86.085117,PhysRevB.90.165138,PhysRevB.99.155143,PhysRevB.104.045134,PhysRevResearch.6.023240,PhysRevResearch.6.013230,reddy2024exploring,PhysRevResearch.2.013191,PhysRevMaterials.2.075003,PhysRevB.75.035122}.

We then explore the trends in the early series of 3$d$ TMOs, specifically from 3$d^1$ to 3$d^3$. It is worth noting that the interaction parameters for the late series of 3$d^4$, 3$d^5$, 4$d^4$, 4$d^5$, and 5$d^5$ obtained from $t_{2g}$-$t_{2g}$ are not considered reliable due to significant hybridization between $t_{2g}$ bands and O-2$p$ bands. High-quality Wannier band projections depend on the choice of the energy window (Table~\ref{table1}). For 3$d^4$/4$d^4$/5$d^5$, an empirical energy window was used, resulting in  $t_{2g}$ Wannier projections that closely match the original DFT bands.

In the early 3$d$ series (3$d^1$-3$d^4$), $V$ fluctuates within a relatively small energy window, ranging from 15.88\,eV (3$d^4$ SrFeO$_3$) to 16.44\,eV (3$d^2$ SrCrO$_3$). Conversely, Wannier orbital spreads exhibit a monotonous increasing tendency from 5.86\,\AA$^2$ (3$d^1$) to 9.15\,\AA$^2$ (3$d^4$). The Wannier orbital spreads in the $t_{2g}$-$t_{2g}$ approximation are generally larger than those in the $d$-$dp$ approximation, suggesting that these correlated orbitals are more delocalized within the $t_{2g}$-$t_{2g}$ approximation. Consequently, $V_{t_{2g}}$ is smaller than $V_{d-dp}$, as shown in Table~\ref{table2} and Table~\ref{table3}.

The fully screened Coulomb interaction $W$ significantly decreases from 3$d^1$ SrVO$_3$ (0.99\,eV) to 3$d^4$ SrFeO$_3$ (0.54\,eV). The reduction of $W_{t_{2g}}$ is more pronounced than $W_{d-dp}$ within the $d$-$dp$ approximation. The constancy of the bare interaction $V$ from 3$d^1$ to 3$d^4$ indicates that the decrease in $W$ is attributed to the increasing full screening effects with the increasing atomic number, consistent with the trends observed in the 3$d$ series within the $d$-$dp$ approximation.
Due to the bandwidth reduction, the distance between unoccupied and occupied states is decreased, contributing to an enhanced total possibility of forming particle-hole excitations and leading to larger polarity and screening effects. Specifically, the $p$-$t_{2g}$ and $t_{2g}$-$e_g$ channels contribute increasingly to the screening from 3$d^1$ SrVO$_3$ to 3$d^4$ SrFeO$_3$ due to the $p$ and $e_g$ Kohn-Sham bands approaching the Fermi level [as shown in the DOSs in Fig.~\ref{Fig-Srband1}(a-d)]. Quantitatively, the ratio $W$/$V$ is about twice larger in 3$d^1$ SrVO$_3$ (0.061) than in 3$d^4$ SrMnO$_3$ (0.033), indicating that the strength of the full screening effect in SrMnO$_3$ is almost twice as strong as in SrVO$_3$ (Table~\ref{table3}).

The partially screened Coulomb interaction $U$ significantly decreases from 3$d^1$ SrVO$_3$ to 3$d^4$ SrFeO$_3$, with the reduction of $U_{t_{2g}}$ being more pronounced than $U_{d-dp}$ within the $d$-$dp$ approximation. While the $t_{2g}$-$t_{2g}$ is eliminated in this case, the screening contribution within $d$-orbitals mainly stems from the $t_{2g}$-$e_g$ channel, which is preserved in $t_{2g}$-$t_{2g}$ while not in $d$-$dp$ approximation. Consequently, $U$ exhibits a similar behavior as $W$ for the early 3$d$ series within the $t_{2g}$-$t_{2g}$ approximation.

In contrast to $J^{d-dp}$, and $J_{bare}^{d-dp}$ within the $d$-$dp$ model, which  all increase with the number of $d$-electrons (the exception is $J_{screened}^{d-dp}$, which exhibits an almost constant value), the $t_{2g}$-$t_{2g}$ Hund's exchange interactions $J_{t_{2g}}$, $J_{bare}$, and $J_{screened}$ within the $t_{2g}$-$t_{2g}$ approximation slightly decrease with the number of $d$-electrons (Table~\ref{table3}). The elimination of $t_{2g}$-$t_{2g}$ transitions in the calculation of $J_{t_{2g}}$ suggests that screening effects induced by $t_{2g}$-$e_g$ transitions are responsible for the decreasing behavior of $J_{t_{2g}}$. The decreasing behavior of $J_{screened}$ indicates that $t_{2g}$-$t_{2g}$ also contributes to the reduction of $J_{screened}$. Despite the apparent influence of $t_{2g}$-$t_{2g}$ screening on $J$, the degree of orbitals localization also plays a role, as evidenced by the monotonous decrease observed in $J_{bare}$ within the $t_{2g}$-$t_{2g}$ Hamiltonian. This suggests that both the screening within $d$-orbital transitions and the localization of orbitals are decisive to $J$, with $J_{t_{2g}}$ being smaller within the $t_{2g}$-$t_{2g}$ Hamiltonian than $J_{d-dp}$ within the $d$-$dp$ Hamiltonian.

\subsection{E. $t_{2g}$-$t_{2g}$ approximation for Sr-based 4$d$ and 5$d$ series}

In the 4$d$ and 5$d$ Sr$B$O$_3$ series, the evolution of DFT bands exhibits similarities with 3$d$ Sr$B$O$_3$ compounds (Fig.~\ref{Fig-Srband2} and \ref{Fig-Srband3}). However, there are notable differences, such as (1) larger crystal field splittings in 4$d$ and 5$d$ compounds, leading to increased $d$-$p$ charge transfer  energy $\Delta(dp)$ and reduced screening in $d$-$p$ channels. (2) The Wannier spreads in 4$d$ and 5$d$ series are generally larger than in 3$d$, indicating more delocalized $t_{2g}$ orbitals. As a result, the Coulomb interactions $W_{t_{2g}}$, $V_{t_{2g}}$, and $U_{t_{2g}}$ are expected to be smaller in 4$d$ and 5$d$ than those in 3$d$ series, varying within a small energy region and exhibiting a ``constant'' behavior. This prediction is supported by the obtained quantities (Table~\ref{table3}), where $U_{t_{2g}}$ in 4$d$ and 5$d$ series shows a general reduction, and it becomes almost constant in 5$d$. This conclusion is supported by Table~\ref{table3} (taking $U$ as example): for the 3$d$ series the $U$ decreases from 3.32\,eV (3$d^1$ SrVO$_3$) to 1.47\,eV (3$d^4$ SrFeO$_3$), for the 4$d$ series the $U$ slightly decreases from 3.20\,eV (4$d^1$ SrNbO$_3$) to 2.19\,eV (4$d^4$ SrRuO$_3$), and for the 5$d$ series the $U$ merely decreases from 3.01\,eV (5$d^1$ SrTaO$_3$) to 2.72\,eV (5$d^4$ SrOsO$_3$).

The $W$ is influenced by a delicate interplay between the full screening effect and the degree of localization of the Wannier $t_{2g}$ orbitals. In addition to the contributions from $d$-$p$ and $t_{2g}$-$e_g$ channel screenings discussed above, the $t_{2g}$-$t_{2g}$ channel screening, which is also influenced by the crystal field splitting, also plays a role. This effect is expected to have a more pronounced impact on $W$ as compared to $U$. Consequently, $W$ exhibits similar trends across the 3$d$, 4$d$, and 5$d$ series: for the 3$d$ series, $W$ decreases from 0.99\,eV (3$d^1$ SrVO$_3$) to 0.54\,eV (3$d^4$ SrFeO$_3$); for the 4$d$ series, $W$ decreases from 0.99\,eV (4$d^1$ SrNbO$_3$) to 0.45\,eV (4$d^4$ SrRuO$_3$); and for the 5$d$ series, $W$ decreases from 0.98\,eV (5$d^1$ SrTaO$_3$) to 0.47\,eV (5$d^4$ SrOsO$_3$).

The $t_{2g}$ orbitals must be more delocalized in 4$d$ and 5$d$ oxides than in 3$d$ oxides since the bare interaction $V$-3$d$ is almost twice as large as that of 4$d$ oxides (Table~\ref{table3}). This observation aligns with the atomic-like basis, where the extension of the 4$d$ atomic wave functions is larger than that of 3$d$ oxides. In the 4$d$ series, $V$ monotonously increases with the rise in $d$-electron numbers, with a reduction of $V$ at 4$d^4$. This is attributed to the hybridization between Ru-$t_{2g}$ and O-2$p$, giving rise to additional screening effect from O-2$p$. Along the lines of the 4$d$ series, the 5$d$ series should also exhibit similar tendency, with $V$ increasing as the $d$-electron numbers increase until $d$-$p$ band overlapping occurs. However, as shown in Fig.~\ref{Fig-SrUJV-t2g}(c), the $V$ of 5$d^1$ SrTaO$_3$ is 10.65\,eV, smaller than that of 5$d^2$ SrWO$_3$ (11.18\,eV) and 5$d^3$ SrReO$_3$ (11.01\,eV). The abnormal behavior in the 5$d$ series is attributed to band hybridization between Sr-$d$ and $B$-5$d$, as illustrated in Fig.~\ref{Fig-Srband3}(a,b), where $d$-bands overlap Sr-$d$ bands. This hybridization influences the degree of orbital localization, thus modifying $V$. Consequently, this hybridization also affects the Wannier spreads and renormalizes the value of $V$.

As indicated by the full screening strength $W$/$V$: within same period, e.g., for 3$d$, the $W$/$V$ becomes smaller as $d$-electron numbers, indicating larger full screening strength. As we discussed above, this is due to the facts that (1) the increased band filling and bandwidth reduction makes the energy distance shorter, (2) more $d$-states become occupied, leading to stronger $d$-$d$ screening, and (3) smaller $d$-$p$ charge transfer energy leads to larger $d$-$p$ screening. Within the same main group, the full screening effects get weaker as indicated by the increasing $W$/$V$: e.g., the $W$/$V$ is 0.061 for 3$d^1$ SrVO$_3$, 0.085 for 4$d^1$ SrNbO$_3$ and 0.091 for 5$d^1$ SrTaO$_3$, which is due to the larger crystal field splitting.

The $U/V$ provides information about the strength of partial screening, primarily from $d$-$p$ screening and other screening contributions (mostly from Sr-$d$ states). One would expect $U/V$ to decrease (indicating increasing partial screening strength) with increasing $d$-electrons within the same group. However, as shown in Table~\ref{table3}, the decrease in $U/V$ for the 4$d$ and 5$d$ series is not as pronounced as in the 3$d$ series. For instance, in the 3$d$ series, $U/V$ decreases from 0.204 (3$d^1$ SrVO$_3$) to 0.092 (3$d^4$ SrFeO$_3$), while in the 5$d$ series, it merely decreases from 0.282 (5$d^1$ SrTaO$_3$) to 0.254 (5$d^4$ SrOsO$3$), remaining almost constant. This phenomenon can be attributed to the slower decrease in $p$-$d$ charge transfer energy and $t_{2g}$-$e_g$ splitting in 4$d$ and 5$d$ series. Within the same main group, the partial screening effects decrease, as indicated by the increasing $U/V$. For example, $U/V$ values are 0.204 for 3$d^1$ SrVO$_3$, 0.277 for 4$d^1$ SrNbO$_3$, and 0.282 for 5$d^1$ SrTaO$_3$. Consequently, on the first hand, $U^{t_{2g}}$ changes significantly in the 3$d$ and 4$d$ series, whereas it remains relatively constant in the 5$d$ series with varying numbers of $d$-electrons (Fig.~\ref{Fig-SrUJV-t2g}). This effect can be attributed to screening, which has a stronger impact on the atomic-like Wannier basis of the 3$d$ series compared to the delocalized nature of the 5$d$ series. On the other hand, the behavior of $W^{t_{2g}}$ is consistent across all 3$d$, 4$d$, and 5$d$ series.

Unlike $V$, $U$, and $W$, the behavior of $J$, $J_{bare}$, and $J_{screened}$ is almost consistent across the 4$d$, and 5$d$ series, with these parameters monotonically decreasing as the number of $d$-electrons increases. However, there is one exception observed in 5$d^1$ SrTaO$_3$, where both $J$ and $J_{bare}$ are almost same as those of 5$d^2$ SrWO$_3$. This deviation can be attributed to the hybridization between Ta-5$d$ and Sr-4$d$, as discussed above. All Hund's exchange $J$ parameters exhibit only minor variations within a narrow energy range.

\subsection{F. $d$-$dp$ or $t_{2g}$-$t_{2g}$ approximations and their scope of applications}

We conducted cRPA calculations using both $t_{2g}$-$t_{2g}$ and $d$-$dp$ approximations to determine Coulomb and Hund’s interaction parameters. It is important to note that the obtained parameters are model-dependent. While direct comparison between the quantities from these two approximations is not feasible, for some cases (as discussed later, with strong $d$-$p$ hybridization), notable differences arise, especially in late TM $AB$O$_3$ materials. For instance, in 3$d^3$ SrMnO$_3$, the $t_{2g}$-$t_{2g}$ approximation yields a Hubbard $U_{t_{2g}}$ of 1.98\,eV, whereas the $d$-$dp$ approximation gives $U_{d-dp}$ =2.79\,eV. This discrepancy arises from the spread of Mn-$d$ characters onto O-ligands due to $d$-$p$ hybridization.

In all 3$d$, 4$d$, and 5$d$ series, as $d$-electron filling increases, O-2$p$ bands approach $B$-$d$ bands, leading to overlapping and reduced $d$-$p$ charge transfer energy $\Delta$($dp$). The increased $d$-$p$ hybridization transfers more $d$ characters to the ligand O sites. Despite the bandwidth reduction, the $t_{2g}$ Wannier orbitals become more delocalized. However, this strong $d$-$p$ hybridization and band overlapping render $t_{2g}$-$t_{2g}$ less appropriate for late TM perovskites. In early TM perovskites, both models yield a valid effective low-energy description, ensuring similar results in many-body calculations. The reliability of the $t_{2g}$-$t_{2g}$ approximation diminishes around the $d^4$ or $d^5$ configuration, as seen in 3$d^4$ SrFeO$_3$, 4$d^4$ SrRuO$_4$ and 5$d^5$ SrIrO$_3$, where non-negligible $t_{2g}$ characters are present even before $t_{2g}$ bands touch O-2$p$ bands [see Fig.~\ref{Fig-Srband1}(d), \ref{Fig-Srband2}(d) and \ref{Fig-Srband3}(e)], there is already non-ignorable $t_{2g}$ characters ranging from $\sim$-6.0\,eV to $\sim$-2.0\,eV. Additionally, the magnetic configuration plays a role; in materials like 3$d^3$ SrMnO$_3$ and 3$d^4$ SrFeO$_3$, high-spin magnetic states result in occupied $e_g$ orbitals, favoring a $d$-$dp$ model for interaction parameter calculations.

The core problem in cRPA calculations within the $t_{2g}$-$t_{2g}$ approximation for $AB$O$_3$ perovskites is the $d$-$p$ bands' overlapping and hybridization. Strong $d$-$p$ hybridization makes it challenging to project onto $t_{2g}$ bands, hindering the obtainment of unique and trustworthy interaction parameters. This issue is effectively addressed with $d$-$dp$ approximations, as they handle $d$-$p$ hybridization in the model constructions.

In the $t_{2g}$-$t_{2g}$ model, Wannier functions are constructed from the threefold-degenerate low-energy $t_{2g}$ bands only. While this procedure projects most $t_{2g}$ character onto the Wannier bands, significant O-2$p$ admixing occurs due to heavy $p$-$d$ hybridization. Stronger $d$-$p$ hybridization results in more ``extended'' $t_{2g}$ Wannier orbitals, showing pronounced tails on the O ligand sites.
In early TM $AB$O$_3$ systems, like 3$d^1$ SrVO$_3$ or 4$d^1$ SrNbO$_3$, the $t_{2g}$ tails are not strongly pronounced due to the large energy gap between TM-$d$ and O-2$p$ states. The $t_{2g}$ character in the energy region from -8.0\,eV to -4.0\,eV, as indicated by the DOS, suggests that the $t_{2g}$-$t_{2g}$ approximation effectively captures the realistic orbital physics of $t_{2g}$ orbitals, making it a suitable approximation for early transition metal $AB$O$_3$ systems.
As the system approaches $d^4$ configurations, exemplified by 3$d^3$ SrMnO$_3$, the $t_{2g}$ bands may be gaped with O-2$p$ bands. However, relying on projected $t_{2g}$ Wannier bands and obtaining interaction parameters for subspace many-body calculations might be questionable when neglecting inter-site interaction terms. If the orbitals are overly extended, significant inter-site interactions may arise, questioning the Hubbard-type model (e.g., in DMFT) with only onsite terms. The divergence between $U_{t_{2g}}$ and $U_{d-dp}$ underscores the importance of including $p$-bands for late transition metal $AB$O$_3$ perovskites to address the $d$-$p$ hybridization issue.

Within the $d$-$dp$ approximation, more precise Wannier projections (for $d$-orbitals) are achieved by constructing projected orbitals from the entire $d$-$p$ group of bands including all $d$-characters. Therefore, applying the $d$-$dp$ approximation for TM $AB$O$_3$ perovskites later than $d^3$ is preferable.

However, a new question emerges regarding the increasing $d$-$p$ hybridization: as it becomes more pronounced, off-diagonal matrix elements related to O-2$p$ orbitals in the interaction parameters matrix, such as $U_{dp}$ or $U_{pp}$, cannot be neglected. Previous research suggests that neglecting $U_{pp}$ might be a reasonable approximation since O-2$p$ states are typically delocalized and fully occupied (except for charge transfer insulator such as cuprate superconductor). Yet, in studies involving $e_g$-orbital materials like LaNiO$_3$ \cite{PhysRevB.90.045128}, $U_{dp}$ terms may be crucial as $e_g$-$p$ interactions/hybridization become stronger than $t_{2g}$-$p$ hybridization.

\section{IV.\,Conclusion}

Determining Coulomb and Hund’s interactions from first-principles for a given low-energy Hamiltonian is a necessary avenue for achieving a truly \emph{ab-initio} description of correlated materials within many-body calculations. Towards this goal, the cRPA method is an elegant approach that enables to obtain the interaction parameters for orbitals within the low-energy correlated subspace. The central idea of the cRPA method is to exclude the contributions within the correlated orbitals when calculating the polarizability within the RPA. The aim to do the constraint is essentially to avoid double counting problem when augmenting the DFT with more sophisticated many-body Hamiltonian approaches, since these methods focus only on the correlated orbitals and already account for the screening within these orbitals. In this regard, it is not meaningful to simply talk about interaction parameters without mentioning the low-energy Hamiltonian.


However, for perovskites TMOs with complicated multi-bands structures, the low-energy Hamiltonian spanning only TM-$t_{2g}$ or entire TM-$d$ states may not be well defined due to the strong hybridization with O-2$p$ states. 
In this situation, the extended O-2$p$ states have to be included in the low-energy Hamiltonian.
Therefore, in the present work, we conducted systematic cRPA calculations of interaction parameters for the TM-$d$ shell of $AB$O$_3$ perovskites ($A$=Sr, Ca; $B$=V to Co, Nb to Rh, Ta to Pt) using two distinct approximations of the low-energy Hamiltonian. The first one is the widely used $t_{2g}$-$t_{2g}$ model, and the other one is the $d$-$dp$ model. For the $t_{2g}$-$t_{2g}$ model, the low-energy Hamiltonian is determined by the TM-$t_{2g}$ orbitals, and the interaction parameters are calculated for the same set of orbitals. For the $d$-$dp$ model, the low-energy Hamiltonian is described by both TM-$d$ and O-$2p$ orbitals, while the interaction parameters are obtained only for the TM-$d$ shell. For both models, Sr$B$O$_3$ and Ca$B$O$_3$ (see Appendix) yield similar trends for the interaction parameters over the occupancy of $d$ orbitals.

Within the $t_{2g}$-$t_{2g}$ framework, the bare  Coulomb interaction $V$ fluctuates within a small energy window (less than 1.0\,eV), which is influenced by an interplay between bandwidth reduction and orbital delocalization with increasing $d$-electron number. This demonstrates the atomic-like behavior of $AB$O$_3$, especially for 3$d$ and 4$d$ series, with $V$ monotonously increasing as $d$-electron number rises. This is because the 3-4$d$ orbitals are more localized than the 5$d$ orbitals and their orbital localization is not affected by Sr-4$d$ hybridization. The partially screened Coulomb interaction $U$ decreases with increasing $d$-electron number due to the enhanced $t_{2g}$-$e_g$ and $d$-$p$ screening, contributing significantly to the rest polarizability. The fully screened Coulomb interaction $W$ also decreases monotonously due to the increasing full screening effects. Hund’s interactions ($J_{bare}$, $J$, and $J_{screened}$) all decrease with rising $d$-electron numbers, implying that Hund’s interactions are mediated by enhanced $d$-orbital delocalization. 

However, for the $d$-$dp$ model, we observed that $U$ tends to exhibit a maximum at $d^4$ (for 3$d$ and 4$d$ series) or $d^5$ (for 5$d$ series). This maximum arises from the competition between orbital localization and $d$-$d$ (and also $d$-$p$) screening effect. The $V$ monotonously increases along the same period, indicating that it is predominantly influenced by orbital localization. The $W$ is influenced by the interplay between the enhancement and reduction of full screening effects, resulting in a minimum at $d^3$ for 3$d$, 4$d$ and 5$d$ perovskites. Both $J$ and $J_{bare}$ show similar tendencies as $V$, suggesting that they are determined by orbital localization, while $J_{screened}$ exhibits a ``decrease-increase'' behavior (with a minimum at $d^3$, similar to $W$), indicating the dominate role of full screening effects in determining $J_{screened}$.

This work demonstrates that the actual value of $U$ depends on the subtle competition between orbital localization and the strength of screenings excluding the contribution from correlated orbitals, which are in turn subject to the specific band structure modulated by the $d$-orbital filling. Moreover, this work highlights the importance of defining appropriate low-energy Hamiltonian, which is directly linked to the physical meaning of the obtained $U$.
The conclusion of this systematic study is general and can be extended to other materials and other physical scenarios.

\subsection{Acknowledgments}

\begin{acknowledgments}

L.~S.~is thankful for the starting funds from Northwest University.
L.~S.~and C.~F.~were supported by the Austrian Science Fund (FWF) SFB project ViCoM (Grant No.~F41).
C.~F. acknowledges the funding by the European Union—Next Generation EU—“PNRR—M4C2, investimento 1.1-Fondo PRIN 2022”—“Superlattices of relativistic oxides”
(ID No. 2022L28H97, CUP D53D23002260006).
P.~L. acknowledges the funding from the National Key R\&D Program of China  (Grant No.~2021YFB3501503), the National Natural Science Foundation of China (Grant No.~52188101), and the Liaoning Province Science and Technology Planning Project (E439LA01).
Calculations have been done on the Vienna Scientific Clusters (VSC).

For the input and output files of the calculation examples discussed in this paper, please refer to 10.17172/NOMAD/2024.07.31-1 \cite{NOMAD}.
\end{acknowledgments}

\section{Appendix}

\subsection{A. Interaction parameters for Ca-based 3$d$, 4$d$ and 5$d$ series}

To corroborate our findings in the discussion of Sr$B$O$_3$ perovskites, we performed electronic structure and interaction parameter calculations for Ca$B$O$_3$ perovskites. The trends observed in Ca$B$O$_3$ are basically consistent with those in Sr$B$O$_3$, reinforcing our earlier conclusions.

\begin{figure*}
\centering
\includegraphics[width=18.0cm]{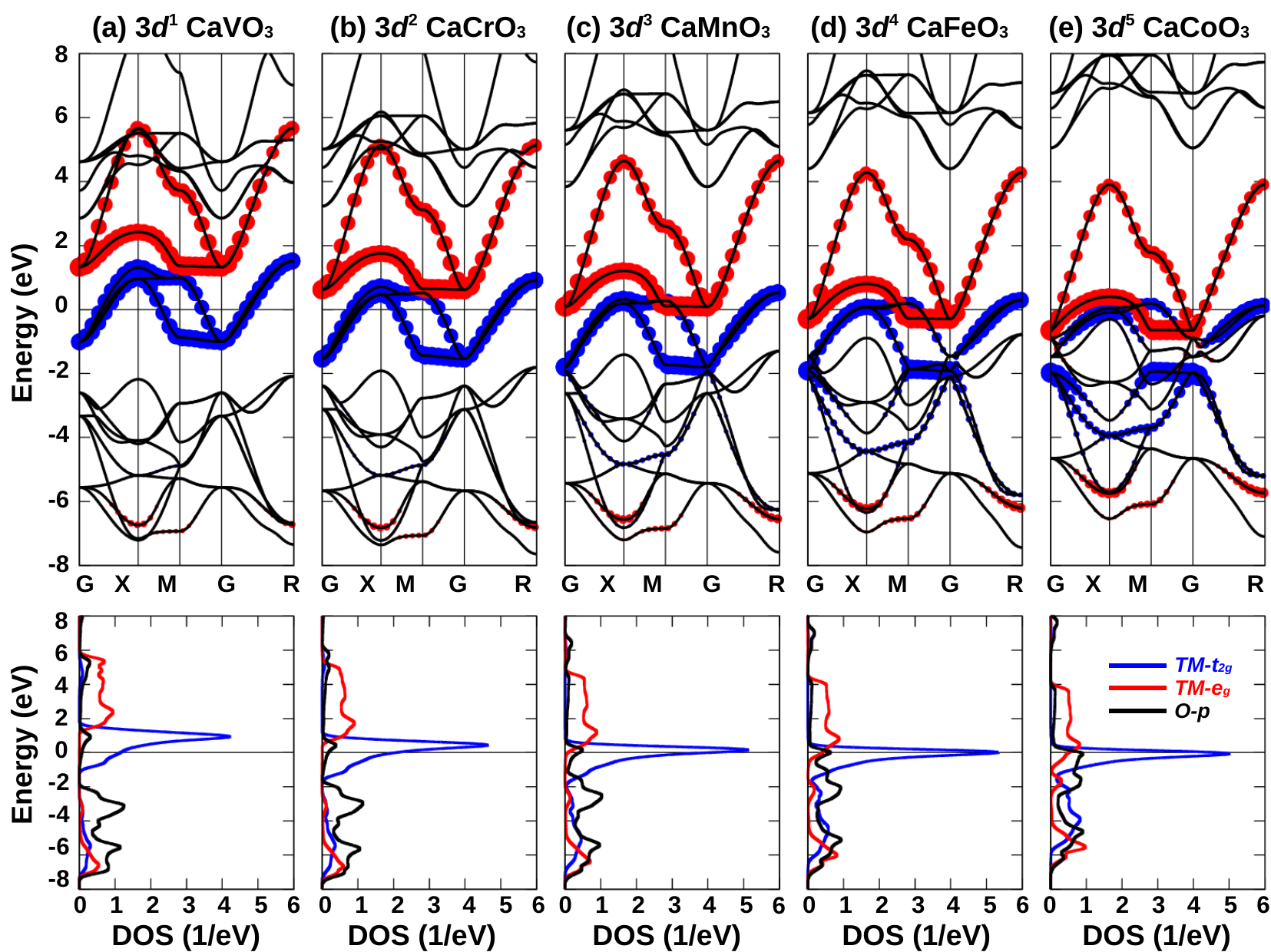}
\caption{(Color online) 
DFT bands (top panels) and density of states (bottom panels)
of Ca$B$O$_3$ ($B$=V, Cr, Mn, Fe, and Co).
The size of the blue and red points indicates contributions from $t_{2g}$ and $e_g$ orbitals.
}
\label{Fig-Caband1}
\end{figure*}

\begin{figure*}
\centering
\includegraphics[width=18.0cm]{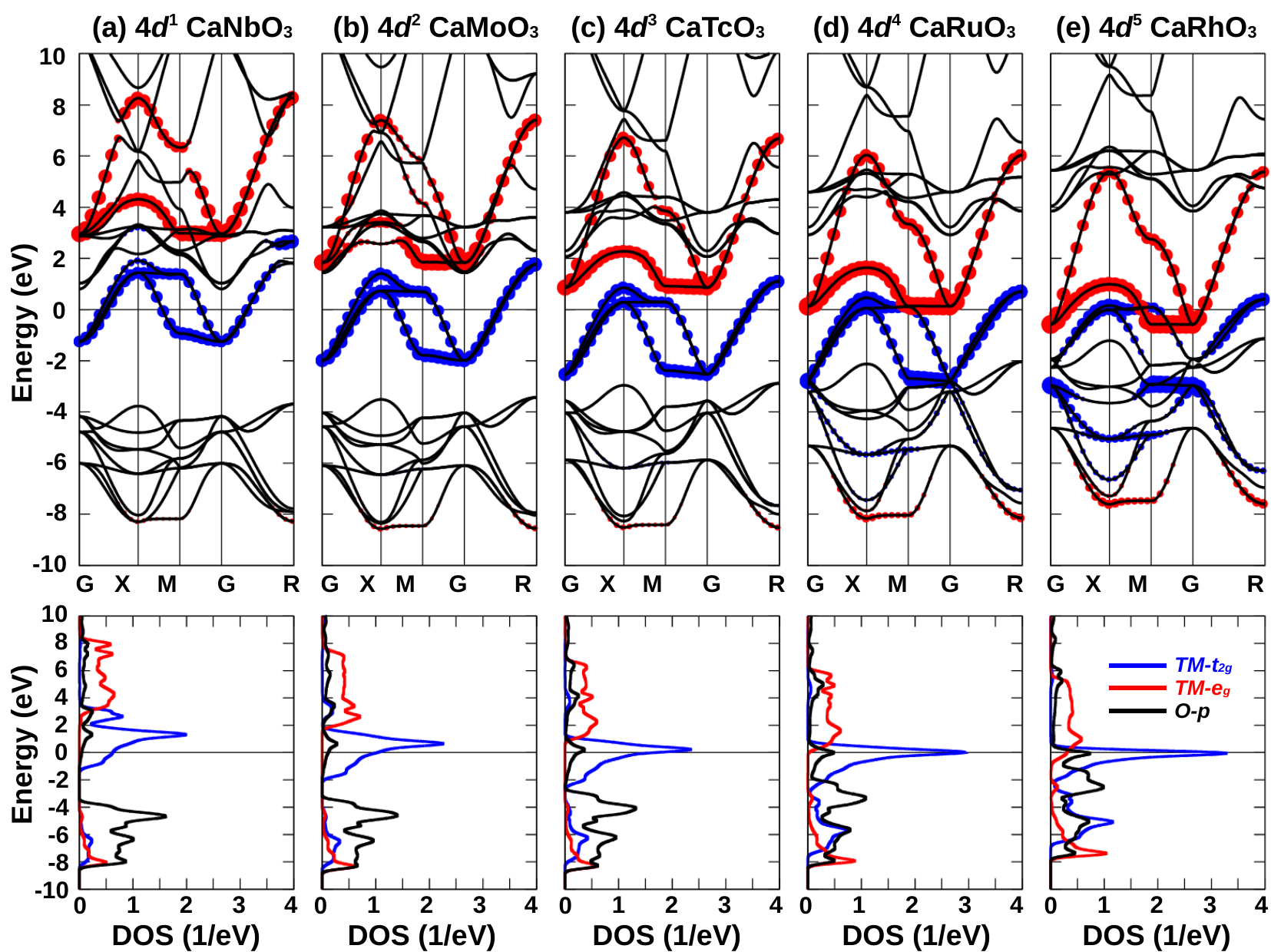}
\caption{(Color online) 
DFT bands (top panels) and density of states (bottom panels)
of Ca$B$O$_3$ ($B$=Nb, Mo, Tc, Ru, and Rh).
The size of the blue and red points indicates contributions from $t_{2g}$ and $e_g$ orbitals.
}
\label{Fig-Caband2}
\end{figure*}

\begin{figure*}
\centering
\includegraphics[width=18.0cm]{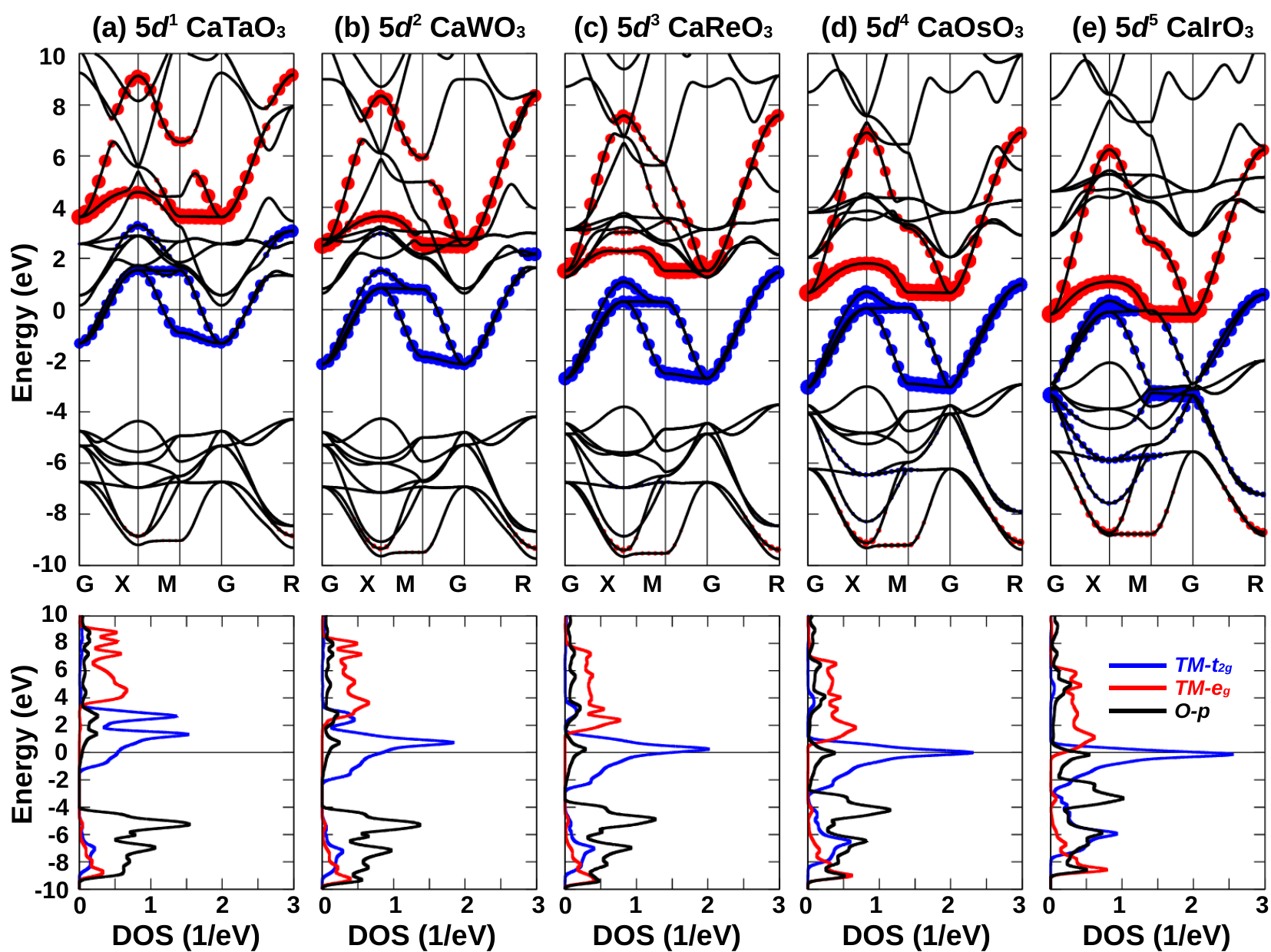}
\caption{(Color online) 
DFT bands (top panels) and density of states (bottom panels)
of Ca$B$O$_3$ ($B$=Ta, W, Re, Os, and Ir).
The size of the blue and red points indicates contributions from $t_{2g}$ and $e_g$ orbitals.
}
\label{Fig-Caband3}
\end{figure*}

\begin{figure*}
\centering
\includegraphics[width=18.0cm]{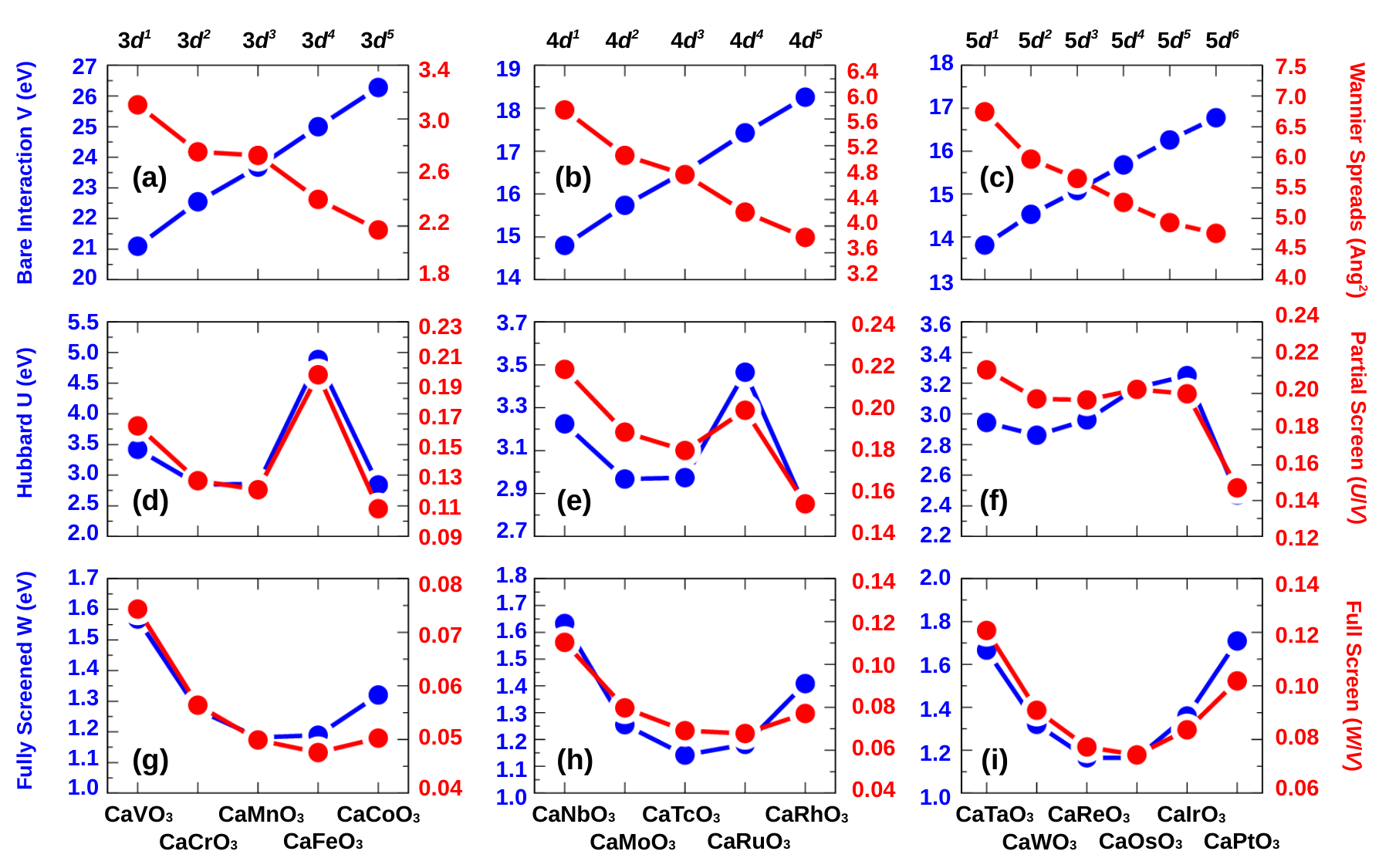}
\caption{Bare interaction $V$ vs. spreads of $d$-orbitals (upper panels), Hubbard interaction $U$ (partially screened interaction) vs. partial screening ($U$/$V$) (middle panels), and fully screened interaction $W$ vs. full screening ($W$/$V$) (bottom panels) of Ca$B$O$_3$, within the $d$-$dp$ approximation.}
\label{Fig-CaUVJ-dp}
\end{figure*}

The band structures of Ca$B$O$_3$ are shown in Fig.~\ref{Fig-Caband1}-\ref{Fig-Caband3}. Results of interaction parameters obtained from the $d$-$dp$ approximation are presented in Fig.~\ref{Fig-CaUVJ-dp} and the right panel of Table~\ref{table3}. In Fig.~\ref{Fig-CaUVJ-dp}(a-c), the $V$ for 3-5$d$ Ca$B$O$_3$ increases as $d$-orbitals become more localized, indicated by decreasing Wannier orbital spreads. $V$ is larger in 3$d$ than in 4$d$ and 5$d$ due to the more localized $d$-orbitals in 3$d$ series. Hubbard $U$ also exhibits a tendency similar to Sr$B$O$_3$: local maxima occur at $d^1$ and $d^4$ for 3$d$ and 4$d$ series, and $d^1$ and $d^5$ for 5$d$ series. Notably, the Hubbard $U$ of 3$d^4$ CaFeO$_3$ (4.89\,eV) is $\sim$1.4\,eV stronger than in 3$d^4$ SrFeO$_3$, suggesting a greater contribution of $d$-$d$ screening in 3$d^4$ CaFeO$_3$ to full screening. However, no significant differences are observed in their electronic bands and DOSs [compare Fig.~\ref{Fig-Srband1}(d) to Fig.~\ref{Fig-Caband1}(d)]. 

\begin{figure*}
\centering
\includegraphics[width=18.0cm]{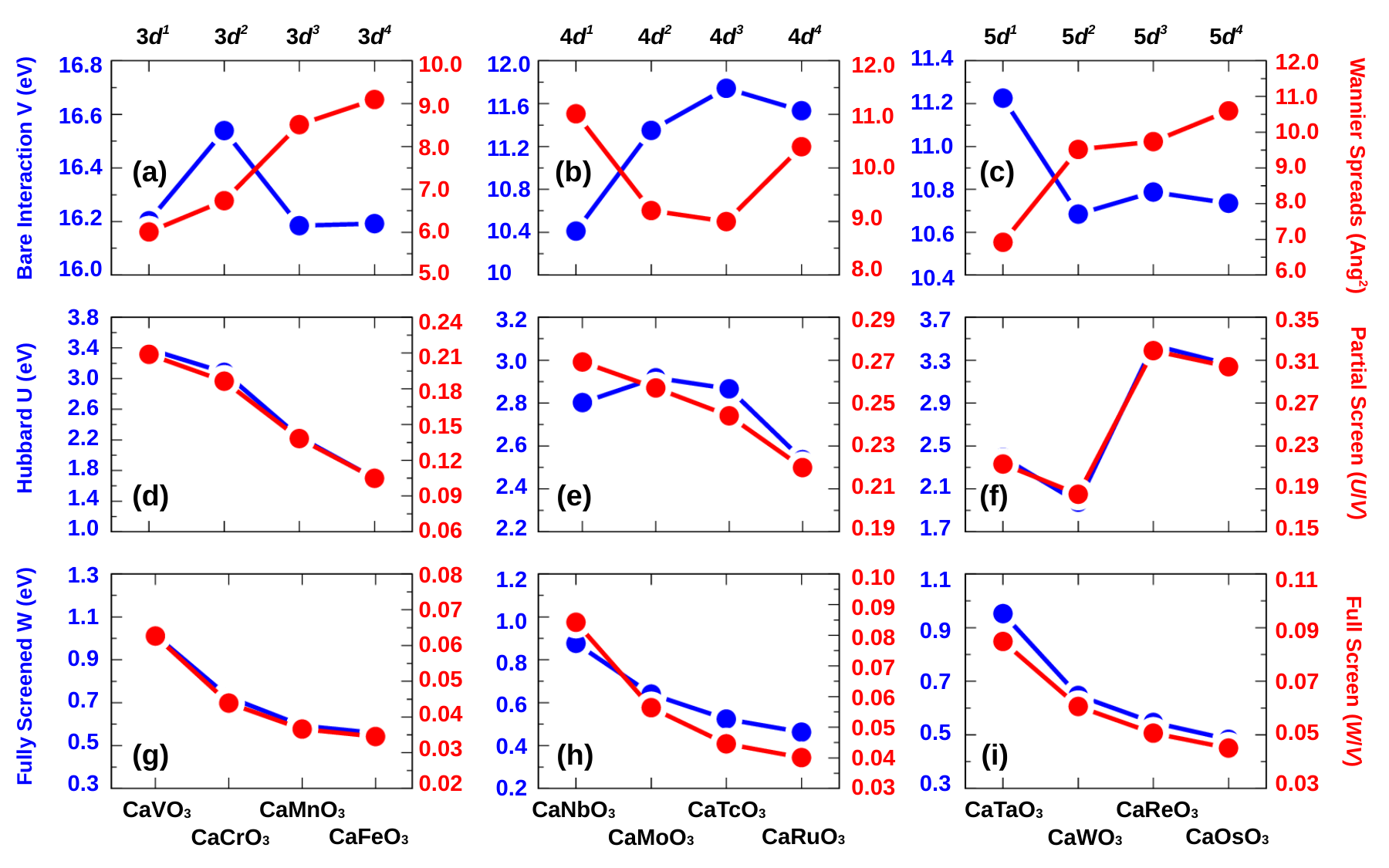}
\caption{Bare interaction $V$ vs. spreads of $d$-orbitals (upper panels), Hubbard interaction $U$ (partially screened interaction) vs. partial screening ($U$/$V$) (middle panels), and fully screened interaction $W$ vs. full screening ($W$/$V$) (bottom panels) of Ca$B$O$_3$, within the $t_{2g}$-$t_{2g}$ approximation.}
\label{Fig-CaUVJ-t2g}
\end{figure*}

Within $t_{2g}$-$t_{2g}$ approximation (Fig.~\ref{Fig-CaUVJ-t2g}), basically all the Hund's exchange interaction parameters of Ca$B$O$_3$ exhibit similar tendencies as in Sr$B$O$_3$. All the bare $V$ merely varies within a small energy window about $\sim$1.0\,eV. In the more extended 4$d$ series the $V$ follows atomic-like behavior: increase as $d$-electron filling for early transition metal perovskites ($d$$\leq$3). For 5$d$ series, the 5$d^1$ configuration is affected by the Ta-5$d$ and Sr-4$d$ hybridization, so both the Wannier localization and the $V$ are modified by this hybridization, leading to the breaking down of the atomic-like behavior. The tendency of fully screened $W$ is same as in Sr-based series: deceases as $d$-electron increases, indicating the full screening effect gets stronger. For both 3$d$ and 4$d$ series of Ca$B$O$_3$, the tendency of Hubbard $U$ is same as in 3$d$ and 4$d$ series of Sr$B$O$_3$, i.e., they decrease as $d$-electron filling increases. However, different tendency is found for the 5$d$ series. The Hubbard $U$ of 5$d^1$ CaTaO$_3$ (2.39\,eV) and 5$d^2$ CaWO$_3$ (1.98\,eV), are remarkably smaller than those in 5$d^3$ CaReO$_3$ (3.44\,eV) and 5$d^4$ CaOsO$_3$ (3.26\,eV). We attribute this to the Sr-4$d$ and TM-5$d$ hybridization as shown in the bands of 5$d^1$ CaTaO$_3$ and 5$d^2$ CaWO$_3$. The $t_{2g}$ bands of CaTaO$_3$ and CaWO$_3$ are heavily overlapped. This enhances the Wannier spreads of projected $t_{2g}$
orbitals and reduced the Hubbard $U$, and most importantly, the $d$-$d$ screening is reduced. Thus the remaining $d$-$p$ screening plays an effecttive role in reducing $V$ to $U$ at both 5$d^1$ and 5$d^2$ configurations.

Within the $t_{2g}$-$t_{2g}$ approximation (Fig.~\ref{Fig-CaUVJ-t2g}), Hund’s exchange $J$ for Ca$B$O$_3$ show similar trends to Sr$B$O$_3$. Bare $V$ varies within a small energy window of $\sim$1.0\,eV. In the more delocalized 4$d$ series, $V$ increases with $d$-electron filling for $d$$\leq$3. For 5$d$ series, the 5$d^1$ configuration is influenced by Ta-5$d$ and Sr-4$d$ hybridization, altering both Wannier localization and hence $V$. This breakdown of atomic-like behavior is evident. The fully screened $W$ follows the same trend as in Sr-based series: decreasing as $d$-electron filling increases, indicating a strengthening full screening effect.
In both 3$d$ and 4$d$ series of Ca$B$O$_3$, Hubbard $U$ decreases with increasing $d$-electron filling, mirroring the trend in 3$d$ and 4$d$ series of Sr$B$O$_3$. However, a different trend is observed for the 5$d$ series. The Hubbard $U$ of 5$d^1$ CaTaO$_3$ (2.39\,eV) and 5$d^2$ CaWO$3$ (1.98\,eV) is significantly smaller than in 5$d^3$ CaReO$_3$ (3.44\,eV) and 5$d^4$ CaOsO$_3$ (3.26 eV). This can be attributed to Sr-4$d$ and $B$-5$d$ hybridization, evident in the bands of 5$d^1$ CaTaO$_3$ and 5$d^2$ CaWO$_3$ (Fig.~\ref{Fig-Caband3}). The $t_{2g}$ bands of CaTaO$_3$ and CaWO$_3$ heavily overlap, enhancing the Wannier spreads of projected $t_{2g}$ orbitals and reducing Hubbard $U$. Importantly, the $d$-$d$ screening is reduced, allowing the remaining $d$-$p$ screening to play a significant role in reducing $V$ to $U$ in both 5$d^1$ and 5$d^2$ configurations.

%

\end{document}